\documentclass[aps,prx,reprint,superscriptaddress]{revtex4-1}

\usepackage{amsmath}
\usepackage{amsfonts}

\usepackage{siunitx}
\usepackage{physics} 

\usepackage{graphicx}

\usepackage{xcolor}

\newcommand{\vect}[1]{\boldsymbol{#1}}

\graphicspath{{./figures/}}

\begin{document}


\title{Hole Spin Qubits in Ge Nanowire Quantum Dots:\\
	Interplay of Orbital Magnetic Field, Strain, and Growth Direction
}


\author{Christoph Adelsberger}
\affiliation{Department of Physics, University of Basel, Klingelbergstrasse 82, 4056 Basel, Switzerland}
\author{M\'onica Benito}
\affiliation{Department of Physics, University of Basel, Klingelbergstrasse 82, 4056 Basel, Switzerland}
\affiliation{Institute of Physics, University of Augsburg, 86135 Augsburg, Germany}
\author{Stefano Bosco}
\affiliation{Department of Physics, University of Basel, Klingelbergstrasse 82, 4056 Basel, Switzerland}
\author{Jelena Klinovaja}
\affiliation{Department of Physics, University of Basel, Klingelbergstrasse 82, 4056 Basel, Switzerland}
\author{Daniel Loss}
\affiliation{Department of Physics, University of Basel, Klingelbergstrasse 82, 4056 Basel, Switzerland}


\date{\today}

\begin{abstract}
Hole spin qubits in quasi one-dimensional structures are a promising platform for quantum information processing because of the strong spin-orbit interaction (SOI). We present analytical results and discuss device designs that optimize the SOI in Ge semiconductors.
We show that at the magnetic field values at which qubits are operated, orbital effects of magnetic fields can strongly affect the response of the spin qubit. We study one-dimensional hole systems in Ge under the influence of electric and magnetic fields applied perpendicularly to the device. In our theoretical description, we include these effects exactly. The orbital effects lead to a strong renormalization of the $g$-factor. 
We find a sweet-spot of the nanowire (NW) $g$-factor where charge noise is strongly suppressed and present an effective low-energy model that captures the dependence of the SOI on the electromagnetic fields.
Moreover, we compare properties of NWs with square and circular cross-sections with ones of gate-defined one-dimensional channels in two-dimensional Ge heterostructures. 
Interestingly, the effective model predicts a flat band ground state for fine-tuned electric and magnetic fields.
By considering a quantum dot (QD) harmonically confined by gates, we demonstrate that the NW $g$-factor sweet spot is retained in the QD.
Our calculations reveal that this sweet spot can be designed to coincide with the maximum of the SOI, yielding highly coherent qubits with large Rabi frequencies. We also study the effective $g$-factor of NWs grown along different high symmetry axes and find that our model derived for isotropic semiconductors is valid for the most relevant growth directions of non-isotropic Ge NWs. Moreover, a NW grown along one of the three main crystallographic axes shows the largest SOI. Our results show that the isotropic approximation is not justified in Ge in all cases. 

\end{abstract}


\maketitle

\section{Introduction}
Important challenges of the scalability of spin qubits
defined in quantum dots (QD)~\cite{Loss1998} can be overcome by making qubits electrically controllable. In the case of electrons,
one can take advantage of the small intrinsic spin-orbit interaction (SOI)~\cite{Nowack2007} or gradients of magnetic field~\cite{Mi2018,Benito2019,Croot2020}. The physics of hole spins in semiconductors such as germanium (Ge) and silicon (Si) 
has attracted much attention lately because it naturally enables stronger SOI than in electron systems~\cite{Kloeffel2018,Kloeffel2011,Froning2021,Hao2010,Hu2011,Scappucci2020,Terrazos2021}.
Another great advantage of hole spins is their tunable response to the hyperfine interactions~\cite{Fischer2010,Maier2012,Klauser2006,Prechtel2016,Testelin2009,Fischer2008}, a major decoherence channel in spin qubits~\cite{Witzel2006,Yao2006,Cywinski2009,Khaetskii2002,Coish2004,Hanson2007}, which can be made far weaker than in electron QDs~\cite{Bosco2021a,Hu2007}. Furthermore in Si and Ge, the hyperfine interactions can also be minimized by isotopically purifying the material~\cite{Itoh1993,Itoh2014}. 
The effective low-dimensional physics of hole systems depends strongly on the details of the confinement, the material strain, and the applied electromagnetic fields~\cite{Bulaev2005,Bulaev2007,Kloeffel2011,Kloeffel2014,Kloeffel2018,Bosco2021,Philippopoulos2020,Michal2021,Hu2007}.
In particular, the strongest SOI, enabling the fastest and most power-efficient operations, is reached in quasi one-dimensional structures, which can be fabricated in different ways~\cite{Kloeffel2018, Bosco2021, Watzinger2018}.

To define spin qubits, a magnetic field is necessary to energetically split different spin states. When a magnetic field is applied perpendicular to the axis of a one-dimensional nanowire (NW), the magnetic orbital effects can strongly influence the performance of the qubit. In two-dimensional heterostructures with a magnetic field applied in-plane, the influence of  these orbital effects strongly depends on the width of the two-dimensional electron or hole gas. In this case one can observe a correction of the $g$-factor~\cite{Stano2018} and a renormalization of the effective mass~\cite{Stano2019} depending on the design of the dot. 

In this work, we analyze the SOI, the effective $g$-factor, and the effective mass of the lowest-energy states in NW with rectangular or circular cross-section and in squeezed planar heterostructures. We compare different designs and fully account for the orbital effects in a moderate magnetic and electric field. 
In contrast to Ref.~\cite{Kloeffel2018}, we use here a different basis which allows us to treat these orbital effects exactly in our analytical calculations.
Interestingly, we predict that, where the SOI is maximal, the $g$-factor can be fine-tuned to be in a sweet spot at which the charge noise is negligible. Similar sweet spots have been predicted in Ref.~\cite{Wang2021} in hole systems possessing a SOI that is cubic in momentum. In these systems, because of the cubic SOI, the Rabi frequency in electric-dipole-induced spin resonance (EDSR) experiments is predicted to be two orders of magnitude smaller than in elongated QDs~\cite{Bosco2021}. 
Furthermore, we find that the effective masses of the low-energy holes depend strongly on both electric and magnetic fields and become spin-dependent at finite magnetic fields. Interestingly, in an extreme case, the lowest energy band can be tuned to be flat. We envision that this flatness potentially could promote hole NWs as a new playground for simulating strongly correlated matter. 
Also, by extending our analysis to include the cubic anisotropies of the valence band, we find that these corrections can strongly affect the $g$-factor for certain growth directions of the NW, especially in the presence of strain.

 \begin{figure}[]
 	\includegraphics{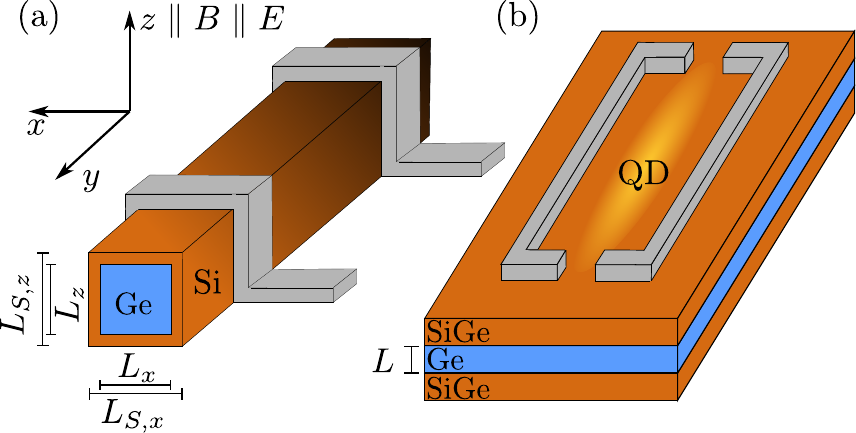}
 	\caption{(a) Sketch of a rectangular Ge NW with side lengths $L_x$ and $L_z$. The NW can be covered by a Si shell which induces strain in the Ge core. The NW including shell has total side lengths $L_{S,x}$ and $L_{S,z}$. The gates (gray) can be used to apply electric fields and to define an elongated QD. (b) Sketch of a planar Ge/SiGe heterostructure with a gate-defined one-dimensional  channel. The height of the Ge layer in the center is $L$ and the channel is electrostatically confined in the $x$-direction with a harmonic confinement length $l_x$. \label{fig:schematic}}
 \end{figure}

 This work is structured as follows. In Sec.~\ref{sec:Model}, we introduce the model of low energy holes.
 In Sec.~\ref{sec:DRSOC}, we focus on a NW with rectangular cross-section and calculate the SOI that is linear in momentum and of the direct Rashba type due to strong heavy hole (HH) light hole (LH) mixing~\cite{Kloeffel2011,Kloeffel2018}. First, we derive  analytic expressions for the dependence of the SOI on external electric fields with and without magnetic orbital effects and compare them with numerical results.
 The complete low-energy model, the effective $g$-factor, and the effective masses are analyzed in Sec.~\ref{sec:geff_meff}. 
We show from a numerical analysis that the $g$-factor is strongly renormalized by orbital effects. Moreover, we find  a spin-dependent effective mass term that  strongly depends on magnetic and electric fields. In Sec.~\ref{sec:Geometries}, we compare NWs with square and circular cross-sections to squeezed dots in Ge/SiGe planar heterostructures, also including strain. 
In Sec.~\ref{sec:QDphys} we analyze the $g$-factor of a QD formed by gate confinement along the NW. In these system, we predict fast Rabi frequencies at low power at electric field values compatible with ones needed for a $g$-factor sweet spot, thus enhancing the performance of the qubit.
We conclude this section by showing that the effective model breaks down for certain fine-tuned electric and magnetic fields, where the lowest band in the NW spectrum becomes flat.
In Sec.~\ref{sec:beyondSA}, we study NWs grown along different typical growth directions, and, in particular, we focus on the interplay of cubic anisotropies and orbital effects. In Sec.~\ref{sec:conclusion} we conclude and summarize our results and details to our calculations are appended.

\section{Model of the nanowire \label{sec:Model}}

The general Hamiltonian describing properties of low-energy holes in diamond-lattice semiconductors is written as
\begin{align}
H = H_\mathrm{LK} + H_\mathrm{BP} + H_Z + H_E + V, \label{eqn:FullHamiltonian}
\end{align}
where $ H_\mathrm{LK}$ is the Luttinger-Kohn (LK) Hamiltonian~\cite{Luttinger1955,Luttinger1956} describing the spin-3/2 holes near the $\Gamma$ point. In addition, $H_\mathrm{BP}$ is the Bir-Pikus (BP) Hamiltonian~\cite{Bir1974} capturing the strain of the lattice, and $H_Z$ is the Zeeman Hamiltonian describing the coupling of the spin to an external magnetic field. The term $H_E$ includes the electric field generated by an externally applied gate potential. In order to define a quasi one-dimensional channel, we consider a confinement potential $V$ that models the cross-section as depicted in Fig.~\ref{fig:schematic}(a) or the harmonic confinement along the $x$-direction produced by gates in a planar Ge/SiGe heterostructure, see Fig.~\ref{fig:schematic}(b).

In Ge, the material dependent LK parameters are $\gamma_1 = 13.35$, $\gamma_2 = 4.25$, and $\gamma_3 = 5.69$~\cite{Lawaetz1971}. Since $(\gamma_3 - \gamma_2)/\gamma_1 \approx \SI{10.8}{\percent} $ is rather small, we will use the isotropic LK Hamiltonian in the following, unless we indicate otherwise. The isotropic LK Hamiltonian $H_\mathrm{LK}$ is written as~\cite{Luttinger1955,Luttinger1956,Kloeffel2018,Bychkov1984,Winkler2008,Lawaetz1971,Lipari1970}
\begin{align}
	&H_\mathrm{LK} = \frac{\hbar^2}{2 m} \left[\gamma_k \vect{k}^2 - 2 \gamma_s \left(\vect{k} \cdot \vect{J}\right)^2 \right]+ H_\mathrm{orb}, \label{eqn:LK_Hamiltonian_sphericalApp}
\end{align}
where the orbital effects of the magnetic field are given by
\begin{align}
	&H_\mathrm{orb}  =\frac{\hbar e}{2 m} \bigg\{\gamma_k \left(\frac{e}{\hbar}\vect{A}^2 + 2\vect{k}\cdot \vect{A}\right) - \frac{2 \gamma_s  e}{\hbar}\left(\vect{A}\cdot\vect{J}\right)^2\nonumber\\
	&-4\gamma_s \left[k_x A_x J_x^2  + \left(\left\{k_x, A_y\right\} + \left\{k_y, A_x\right\}\right) \left\{J_x, J_y\right\}+ \mathrm{c.p.}\right]\bigg\}
\end{align}
where $\gamma_2$ and $\gamma_3$ have been replaced by the effective parameter $\gamma_s \approx \num{4.84}$~\footnote{By trying to solve the LK Hamiltonian for arbitrary growth direction of the NW and averaging over the rotation angle it turns out that the choice $\gamma_s = 4.25 \sqrt{1-\frac{3}{8}\left(1-\left(\frac{\gamma_3}{\gamma_2}\right)^2\right)} = 4.84$ is most natural.}, $\gamma_k = \gamma_1 + 5\gamma_s/2$. Here, $\vect{J}$ is the vector of standard spin $3/2$ matrices, $\{A, B\} = (A B + B A)/2$ is the symmetrized anti-commutator, and by ``c.p.'' we mean cyclic permutations. The orbital effects come from the kinematic momentum operator~\cite{Luttinger1956}
\begin{align}
	\vect{\pi} = \vect{k} + \frac{e}{\hbar} \vect{A}, \label{eqn:kinElMom}
\end{align} 
with the canonical momentum operator $k_j = -i \hbar \partial_j$, $j=x,y,z$, the positive elementary charge $e$, and the vector potential $\vect{A}$, which is related to the magnetic field by $\vect{B} = \nabla \times \vect{A}$. In the isotropic approximation, the LK Hamiltonian does not depend on the orientation of the crystallographic axes. We point out that although  in this work our analysis is explicitly restricted to Ge, our analytical results are valid more generally also for holes in GaAs, InAs, and InSb where the isotropic approximation is  applicable~\cite{Winkler2003}. 

In Ge/Si core/shell NWs, the BP Hamiltonian is well-approximated by~\cite{Bir1974,Kloeffel2018}  
\begin{align}
	H_\mathrm{BP}^\mathrm{NW} &= \abs{b} \varepsilon_s  J_{y}^2, \label{eqn:BPHam}
\end{align} 
where $b=\SI{-2.5}{\electronvolt}$~\cite{Bir1974} is the material dependent deformation potential and $\varepsilon_s = \varepsilon_{\perp}- \varepsilon_{yy}$. The strain tensor elements $\varepsilon_{\perp}$ and $\varepsilon_{yy}$ arising from the lattice mismatch between Si and Ge can be estimated following Ref.~\cite{Kloeffel2014} by assuming an homogeneous strain in the core of the NW. With this assumption, the strain tensor elements depend only on the relative shell thickness~\cite{Kloeffel2014, Menendez2010}, and in a Ge core with a Si shell of relative shell thickness $\gamma = (L_{S,x}-L_{x})/L_x=0.1$, they are given by $\varepsilon_{\perp} = \num{-2.1e-3}$ and $\varepsilon_{yy}= \num{-8.3e-3}$. In this case,  the total strain energy is $\SI{0.62}{\percent} \times \abs{b} = \SI{15.5}{\milli\electronvolt}>0.$ In contrast, in a gate-defined one-dimensional channel, see Fig.~\ref{fig:schematic}(b), the strain induced in the Ge layer by the lattice mismatch to the SiGe top and bottom layers  is described by the BP Hamiltonian~\cite{Wang2021,Terrazos2021} 
\begin{align}
	H_\mathrm{BP}^\mathrm{ch} &= \abs{b} \varepsilon_s  J_{z}^2.\label{eqn:BPHam_2d}
\end{align} 
Note that, in contrast to the NW, where the strain tends to favour LHs aligned perpendicular to the NW axis,  here $\abs{b}\varepsilon_s <0$, meaning that the ground state tends to comprise HHs aligned perpendicular to the substrate. For our strain analysis we choose $\abs{\varepsilon_s}= 0.62\%$ as in the NW setup, a value measured in typical Ge heterostructures~\cite{Sammak2019}. 

In addition to the vector potential entering the momentum operators in Eq.~\eqref{eqn:kinElMom}, an external magnetic field $\vect{B}$ couples directly to the spin degree of freedom via the Zeeman Hamiltonian
\begin{align}
H_Z = 2 \kappa \mu_B \vect{B}\cdot \vect{J} ,
 \label{eqn:HamZeeman}
\end{align}
where $\kappa$ is a material dependent parameter (for Ge $\kappa = 3.41$).
We neglect here the anisotropic Zeeman term, which is less relevant in NWs~\cite{Luttinger1956,Lawaetz1971}. 

Finally we include a homogeneous electric field via the Hamiltonian
\begin{align}
H_E = -e \vect{E} \cdot \vect{r}. \label{eqn:HamElectric}
\end{align}
When the hole wavefunction is strongly confined in two directions and there is an inversion symmetry breaking electric field, the system presents a strong so-called direct Rashba SOI~\cite{Kloeffel2011,Kloeffel2018} that is linear in momentum and is important for fast all-electric manipulation of spin qubits via EDSR.

\section{Direct Rashba spin-orbit coupling \label{sec:DRSOC}}

In this section we calculate the strength of the effective direct Rashba SOI  induced by an electric field applied parallel to the magnetic field (in our case, in $z$-direction) and perpendicular to the NW or quasi one-dimensional structure, which extends in $y$-direction, see Fig. \ref{fig:schematic}. Here, we first consider an infinitely long NW and later, in Sec.~\ref{sec:QDphys}, we confine a QD by a harmonic potential. Since we choose the Landau gauge $\vect{A}=(0,x,0)B$ for $\vect{B}\parallel z$, the translational invariance in $y$-direction is preserved and the canonical momentum $k_y$ is a good quantum number.

\subsection{Model}
First, we assume hard-wall (HW) confinement
\begin{align}
	V(x, z) = 
	\begin{cases}
		0, &|x| < L_x/2\ \mathrm{ and }\ |z| < L_z/2,\\ 
		\infty, &\mathrm{otherwise,}
	\end{cases} \label{eqn:HW_BC}
\end{align}
which describes well a rectangular NW of the width  $L_z$ ($L_x$) in $z$ ($x$)-direction, see Fig.~\ref{fig:schematic}(a). 
We divide the isotropic LK Hamiltonian in Eq.~\eqref{eqn:LK_Hamiltonian_sphericalApp} into three parts,
\begin{align}
	H_\mathrm{LK}^s = H_{zz}+ H_\mathrm{mix} + H_{xy}, \label{eqn:Factorization}
\end{align}
where the addends are defined as
\begin{align}
	H_{zz}= \frac{\hbar^2}{2 m}\left(
	\begin{array}{@{}cccc@{}}
		\gamma_z^\mathrm{H}\pi_z^2 & 0 & 0&0 \\
		0 & \gamma_z^\mathrm{L}\pi_z^2&  0 & 0 \\
		0 & 0 & \gamma_z^\mathrm{L}\pi_z^2&  0  \\
		0 &0 &  0 &\gamma_z^\mathrm{H}\pi_z^2
	\end{array}
	\right),\label{eqn:Hz}
\end{align}
\begin{widetext}
\begin{align}
	&H_\mathrm{mix}  =\frac{\hbar^2 \sqrt{3}\gamma_s}{m}\left[\left(
	\begin{array}{@{}cccc@{}}
		0 & - \{\pi_z, \pi_x\} & 0&0 \\
		- \{\pi_z, \pi_x\}& 0&  0 & 0 \\
		0 & 0 &0&  \{\pi_z, \pi_x\}  \\
		0 &0 &  \{\pi_z, \pi_x\}&0
	\end{array}\right)+i\left(\begin{array}{@{}cccc@{}}
		0 &  \{\pi_z, \pi_y\}& 0&0 \\
		- \{\pi_z, \pi_y\}& 0&  0 & 0 \\
		0 & 0 &0&  - \{\pi_z, \pi_y\}  \\
		0 &0 &   \{\pi_z, \pi_y\}&0
	\end{array}
	\right)\right], \label{eqn:Hint}\\
		&H_{xy} = \frac{\hbar^2}{2m}\left(
		\begin{array}{@{}cccc@{}}
			\gamma_+\{\pi_+,\pi_-\}& 0 & -\sqrt{3}\gamma_s\pi_-^2&0 \\
			0 &\gamma_-\{\pi_+,\pi_-\}&  0 & -\sqrt{3}\gamma_s\pi_-^2 \\
			-\sqrt{3}\gamma_s\pi_+^2 & 0 & \gamma_-\{\pi_+,\pi_-\}&  0  \\
			0 &-\sqrt{3}\gamma_s\pi_+^2  &  0 &\gamma_+\{\pi_+,\pi_-\}
		\end{array}
		\right), \label{eqn:Hxy}
	\end{align}
\end{widetext}
with $\gamma_z^\mathrm{H,L}=\gamma_1\mp 2\gamma_s$, $\gamma_\pm=\gamma_1\pm \gamma_s$, $\pi_\pm = \pi_x\pm i \pi_y$.
Here, $\pi_x$, $\pi_y$, and $\pi_z$ are the components  of the kinematic
momentum as defined in Eq.~\eqref{eqn:kinElMom}. The term 
 $H_{zz}$,
is diagonal in the chosen spin basis $\{+3/2, +1/2, -1/2, -3/2\}$, whereas $H_{xy}$ couples the spins $\pm 3/2$ to the spins $\mp 1/2$, and $H_\mathrm{mix} $ spin $\pm 3/2$ to spin $\pm 1/2$.
In addition we include the Zeeman term $H_Z$ defined in  Eq.~\eqref{eqn:HamZeeman} and the electric field term $H_E$ defined in  Eq.~\eqref{eqn:HamElectric}. We assume that the electric field $\vect{E}$ is applied along the $z$-axis (parallel to the $\vect{B}$-field), and, for convenience, express $H_E$ in terms of the electric length $l_E = [\hbar^2/(2m eE)]^{1/3}$.

\subsection{One-dimensional basis states}
Let us now neglect orbital effects of the magnetic field by assuming $\vect{A} = 0$, hence $\pi_j = k_j, j = x, y, z$.  We  first focus on  the $x$-direction and as in our model there are no fields in  this direction, we only consider 
\small
\begin{align}
	H_{xy}^{k_y=0} = \frac{\hbar^2}{2 m}\left(
	\begin{array}{@{}cccc@{}}
		\gamma_+ k_x^2& 0 & -\sqrt{3}\gamma_s k_x^2&0 \\
		0 &\gamma_- k_x^2&  0 & -\sqrt{3}\gamma_s k_x^2 \\
		-\sqrt{3}\gamma_s k_x^2& 0 & \gamma_- k_x^2&  0  \\
		0 &-\sqrt{3}\gamma_s k_x^2 &  0 &\gamma_+ k_x^2
	\end{array}
	\right) \label{eqn:Hxy_ky0}
\end{align}
\normalsize
in this direction and we choose the basis states
\begin{align}
	\phi_{n_x}^0(x) = \frac{\sqrt{2}}{\sqrt{L_x}} \sin\left[\frac{\pi n_x}{L_x} \left(x+\frac{L_x}{2}\right)\right], \label{eqn:diagBasis_Bperpx}
\end{align}
which respect the HW boundary conditions along the $x$-direction given in Eq.~\eqref{eqn:HW_BC}. We introduce the quantum number $n_x=1,2,\dots$ The off-diagonal matrix elements of the Hamiltonian in Eq.~\eqref{eqn:Hxy_ky0} lead in this basis to superpositions between spin $\pm 3/2$ and $\mp 1/2$.

Next, we consider the $z$-direction and obtain the eigenstates of the Hamiltonian $H_{zz}+H_Z+H_E$. In the absence of electric fields  the eigenfunctions are
\begin{align}
	\phi_{n_z}^0(z) = \frac{\sqrt{2}}{\sqrt{L_z}} \sin\left[\frac{\pi n_z}{L_z} \left(z+\frac{L_z}{2}\right)\right], \label{eqn:diagBasis_Bperpz}
\end{align}
which again respect the HW boundary conditions along $z$ given in Eq.~\eqref{eqn:HW_BC}. Here, we also introduce the quantum number $n_z=1,2,\dots$ The corresponding energy levels, including the Zeeman energy coming from the magnetic field $\vect{B}\parallel z$, are given by
\begin{align}
	\varepsilon^{0,\pm3/2}_z(n_z) &= \frac{\hbar^2 \gamma_z^\mathrm{H}}{2 m} \left(\frac{\pi n_z}{L_z}\right)^2\pm 3 \kappa \mu_B B, \label{eqn:energyZ}\\
	\varepsilon^{0,\pm1/2}_z(n_z) &= \frac{\hbar^2 \gamma_z^\mathrm{L}}{2 m} \left(\frac{\pi n_z}{L_z}\right)^2\pm \kappa \mu_B B.
\end{align}

In contrast, when an electric field $\vect{E}$ is applied along the $z$-axis, the  eigenfunctions can be written in terms of Airy functions,
\begin{align}
	\phi^\mathrm{H(L)}(z) = a\, \text{Ai}\left[g^\mathrm{H(L)}(z)\right]+b\, \text{Bi}\left[g^\mathrm{H(L)}(z)\right], \label{eqn:diagBasis_Bperpz_E}
\end{align}
with 
\begin{align}
g^\mathrm{H(L)}(z)=(\gamma_z^\mathrm{H(L)})^{-1/3}\left(-\frac{z}{l_E}-\frac{2m}{\hbar^2} \varepsilon^\mathrm{H(L)} l_E^2\right). \label{eqn:g}
\end{align}
The values of $\varepsilon^\mathrm{H(L)}(n_z)$ and the coefficients $a$ and $b$  are found numerically by imposing the HW boundary conditions. 
The lowest-energy solutions $\varepsilon^\mathrm{H(L)}(n_z)$, obtained by solving the equation $\phi^\mathrm{H(L)}(0) = \phi^\mathrm{H(L)}(L_z)$ numerically,  are shown in Fig.~\ref{fig:z_energies_L22}(a).

As a result, the total low-energy spectrum of $H_{zz}+H_Z+H_E$ reads:
\begin{align}
	\varepsilon^{\pm3/2}_z(n_z) &= \varepsilon^\mathrm{H}(n_z)\pm 3 \kappa\mu_B B, \label{eqn:energyZ_E}\\
	\varepsilon^{\pm1/2}_z(n_z) &= \varepsilon^\mathrm{L}(n_z)\pm\kappa\mu_B B.
\end{align}
The index $n_z$ labels the solutions (from low to high energy). Importantly, in this case the wave function for spin $\pm 3/2$ is not the same as for spin $\pm 1/2$ [note the factor $\gamma_z^\mathrm{H(L)}$ in Eq.~\eqref{eqn:g}]. This implies, as it will become clear below, that the total wave function cannot be factorized into $z$ and $x,y$ components. However, for low electric fields $\langle \phi^\mathrm{H}_{n'_z\neq n_z}|\phi^\mathrm{L}_{n_z}  \rangle \ll \langle \phi^\mathrm{H}_{n_z}|\phi^\mathrm{L}_{n_z}  \rangle $ and, thus, this factorization is a good approximation [see Figs.~\ref{fig:z_energies_L22}(b) and~\ref{fig:z_energies_L22}(c) below]. 

\subsubsection{Weak electric field approximation \label{sec:weakField}}
 To estimate the weak electric field condition, for simplicity, we  use  variational solutions instead of the numerically exact Airy function solutions. We correct the wave functions corresponding to zero electric field with an exponential factor to write down the  ansatz~\cite{Wang2021},
\begin{align}
	\phi^\mathrm{H(L)}_{n_z}(z) \propto \phi^{0}_{n_z}(z) \exp\left\{-\rho_{n_z}^\mathrm{H(L)}   \left(\frac{z}{L_z}+\frac{1}{2}\right) \right\} , \label{eqn:diagBasis_Bperpz_E_approx}
\end{align}
where the variational parameters $\rho_{n_z}^\mathrm{H(L)} $ minimize the energy of the subsequently orthogonalized  states. 

\begin{figure*}[htb]
	\includegraphics{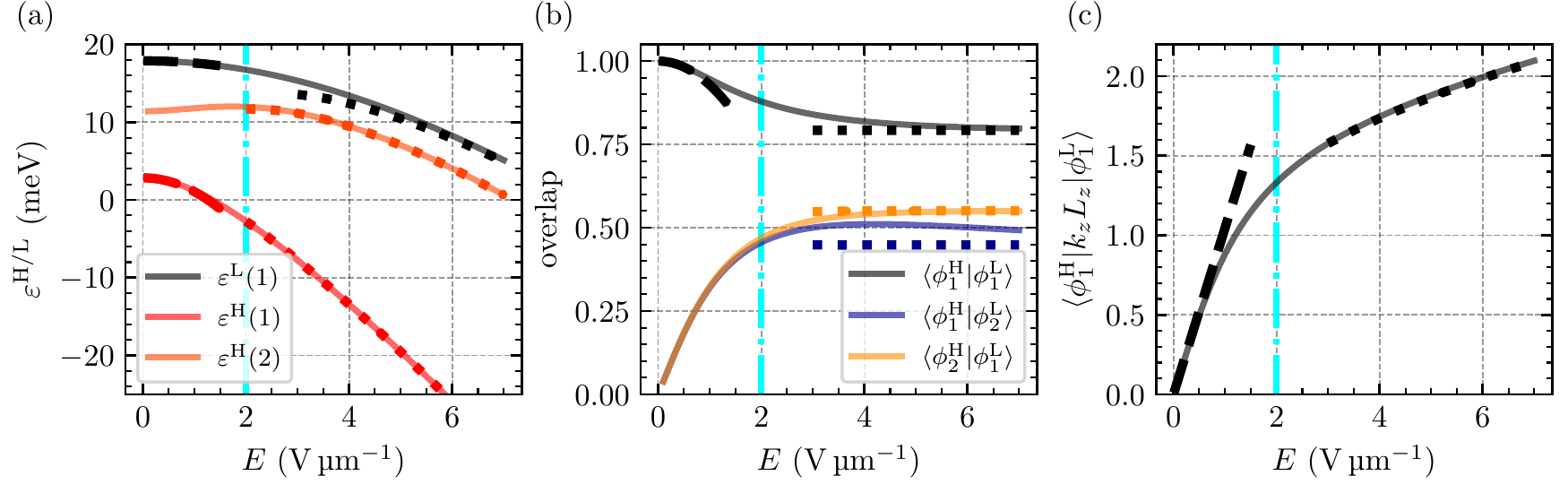}
	\caption{Solutions of  $H_{zz}+H_E$ for the one-dimensional problem in $z$-direction. In all plots the solid lines are the numerically exact Airy function solutions. The dash-dotted cyan line at $E=\SI{2}{\volt\per\micro\meter}$ marks the transition from the weak  to strong electric field regime. (a) Lowest-energy solutions for $\varepsilon^\mathrm{H(L)}(n_z)$ [independent of $B$] as function of electric field $E$ applied in $z$-direction (perpendicular to the NW axis). The dashed lines correspond to the analytical solution defined by Eq.~\eqref{eq:energyZ_E_approx_low} (weak electric field) and the dotted lines to the one defined by  Eq.~\eqref{eq:energyZ_E_approx_high2} (strong electric field).
		(b)  The overlap between wave functions of  HH and LH states. The dashed line corresponds to the analytical solution given by Eq.~\eqref{eq:overlap_approx_low} (weak electric field). The dotted lines represent the result obtained using the wave functions defined in Eq.~\eqref{eq:wavefunctionZ_E_approx_high2}. For weak electric field  $\langle \phi^\mathrm{H}_{n'_z\neq n_z}|\phi^\mathrm{L}_{n_z}  \rangle \ll \langle \phi^\mathrm{H}_{n_z}|\phi^\mathrm{L}_{n_z}  \rangle $. (c) The matrix elements of $k_z$ (in dimensionless units) between the lowest-energy HH and LH states. The dashed line (weak electric field) corresponds to the analytical solution given by  Eq.~\eqref{eq:kzoverlap_approx_low}, and the dotted line (strong electric field)  represents the result  obtained using the wave functions in Eq.~\eqref{eq:wavefunctionZ_E_approx_high2}. In all plots we use $L_z=22\,\text{nm}$.
		Generally, we find that both analytical approximations agree well with exact numerical results shown by the solid lines. These results are used in Sec.~\ref{sec:WithoutOrbEff} to estimate the $g$-factor in Eq.~\eqref{eqn:WOOrbEffLowFieldg} and the SOI in Eq.~\eqref{eq:simple_formula} in the weak electric field case. \label{fig:z_energies_L22}}
\end{figure*}

If the electric field is weak, meaning $L_z/\pi \ll \gamma_1^{1/3} l_E$,  the minimal ground state energy ($n_z = 1$)  is found at
\begin{align}
\rho_1^\mathrm{H(L)}=\frac{(L_z/l_E)^3}{12\pi^2\left(\gamma_z^\mathrm{H(L)}\right)^2}\left(\pi^2-6\right),
\end{align}
and the corresponding energy is
\begin{align}
	\varepsilon^\mathrm{H(L)}(1) &\approx \frac{\hbar^2}{m}\left( \frac{\gamma_z^\mathrm{H(L)}\pi^2}{2L_z^2}-\frac{L_z^4  \left(\pi^2-6\right)^2}{288 \pi^4 \gamma_z^\mathrm{H(L)} l_E^6}+\mathcal{O}\left(\frac{L_z^{10}}{l_E^{12}}\right)\right). \label{eq:energyZ_E_approx_low}
\end{align}
We also give the expression for the overlap 
\begin{align}
\langle \phi^\mathrm{H}_{1}|\phi^\mathrm{L}_{1}  \rangle\approx 1-\frac{(L_z/l_E)^6 \left(\pi^2-6\right)^3 \gamma_s^2}{216 \pi^6(\gamma_1^2-4\gamma_s^2)^2} +\mathcal{O}\left(\frac{L_z^{12}}{l_E^{12}}\right). \label{eq:overlap_approx_low}
\end{align}
and for the matrix elements of $k_z$
\begin{align}
\langle \phi^\mathrm{H}_{1}|k_z|\phi^\mathrm{L}_{1}  \rangle \approx -i \frac{L_z^2  \left(\pi^2-6\right) \gamma_s}{6 l_E^3 \pi^2 (\gamma_1^2-4\gamma_s^2)}+\mathcal{O}\left(\frac{L_z^{8}}{l_E^{9}}\right). \label{eq:kzoverlap_approx_low}
\end{align}

The dashed lines in Fig.~\ref{fig:z_energies_L22}(a), only shown for low electric field, correspond to Eq.~\eqref{eq:energyZ_E_approx_low}. In analogy, we show the low-field approximation for the overlap amplitude
 between the HH and LH wave functions in Fig.~\ref{fig:z_energies_L22}(b). This overlap is important to estimate the SOI.
Finally, the matrix elements of $k_z$ between HH and LH states are also relevant, see Fig.~\ref{fig:z_energies_L22}(c), and we show that the linear approximation used to derive Eq.~\eqref{eq:kzoverlap_approx_low} works well for weak electric field.

\subsubsection{Strong electric field approximation \label{sec:strongField}}

In the opposite limit of strong electric field, the wave function is strongly compressed to the edge. Thus, we approximate the solution by just one Airy function, $\text{Ai}(z)$. In this case, the energy spectrum reads
\begin{align}
	\varepsilon^\mathrm{H(L)}(n_z) &\approx -\frac{\hbar^2}{m}\left(\frac{L_z}{4 l_E^3}+\frac{\left[\gamma^\mathrm{H(L)}\right]^{1/3}}{2 l_E^2}\text{Ai0}(n_z)\right)\label{eq:energyZ_E_approx_high2}
\end{align}
and is shown by a dotted line in Fig.~\ref{fig:z_energies_L22}(a). 
The corresponding wave functions is
\begin{align}
	\phi^\mathrm{H(L)}_{n_z} =&\frac{\text{Ai}\left[\frac{L_z-2z}{2l_E (\gamma^\mathrm{H(L)})^{1/3}}+\text{Ai0}(n_z)\right]}{\sqrt{l_E (\gamma^\mathrm{H(L)})^{1/3}} \text{Ai}'(\text{Ai0}(n_z))}  \ ,\label{eq:wavefunctionZ_E_approx_high2}
\end{align}
with the $n_z^\mathrm{th}$ zero of the Airy function denoted by $\text{Ai0}(n_z)$.
The overlap between  wave functions of  HH and LH states as well as  matrix elements of $k_z$ are shown with dotted lines in  Figs.~\ref{fig:z_energies_L22}(b) and (c). Again, we find an excellent agreement with numerical results in the strong electric field regime.

\subsection{Solution without orbital effects \label{sec:WithoutOrbEff}}

In the following, we use the one-dimensional wave functions derived above to study the SOI. At low magnetic fields, the orbital effects are not expected to give a large contribution to the SOI~\cite{Bosco2021}. Thus, in this subsection we neglect them and calculate a simple formula for the SOI amplitude. 
First, we extend the solution of the Hamiltonian $H_{xy}^{k_y=0}$ at $k_y=0$ [see Eq.~\eqref{eqn:Hxy_ky0}] by including the parity-mixing term 
\begin{align} 
	\frac{H_\mathrm{mix} ^{k_y=0} }{\sqrt{3} \gamma_s} = \frac{\hbar^2}{m}
	\begin{pmatrix}
		0 &  -k_x & 0 &0\\
		-k_x& 0 &0 &0\\
		0 & 0 & 0 & k_x\\
		0 & 0 & k_x & 0
	\end{pmatrix} k_z. \label{eqn:HintBperp}
\end{align}
Later, we will use perturbation theory to include the terms linear in $k_y$:
	\begin{align}
		\frac{H_{k_y}}{  \sqrt{3} \gamma_s} = \frac{\hbar^2}{m} \left(
		\begin{array}{@{}cccc@{}}
			0 & i k_z &  i k_x&0 \\
			-i  k_z &0&  0 &  i k_x \\
			- i k_x& 0 & 0& - i  k_z  \\
			0 &- i k_x&  i  k_z &0
		\end{array}
		\right) k_y. \label{eq:linear_no_orb}
	\end{align}
A good basis of wave functions  is given by
\begin{align}
	\psi^\mathrm{H(L)}_{n_x,n_z}(x,z) =\phi^\mathrm{H(L)}_{n_z}(z) \phi^{0}_{n_x}(x),  \label{eqn:xz_Bperpz_E}
\end{align}
where the  functions $\phi^\mathrm{H(L)}_{n_z}(z)$  and $\phi^{0}_{n_x}(x)$ are introduced in Eqs.~\eqref{eqn:diagBasis_Bperpz_E} and~\eqref{eqn:diagBasis_Bperpx}, respectively.

\begin{figure}[htb]
	\includegraphics[width=\columnwidth]{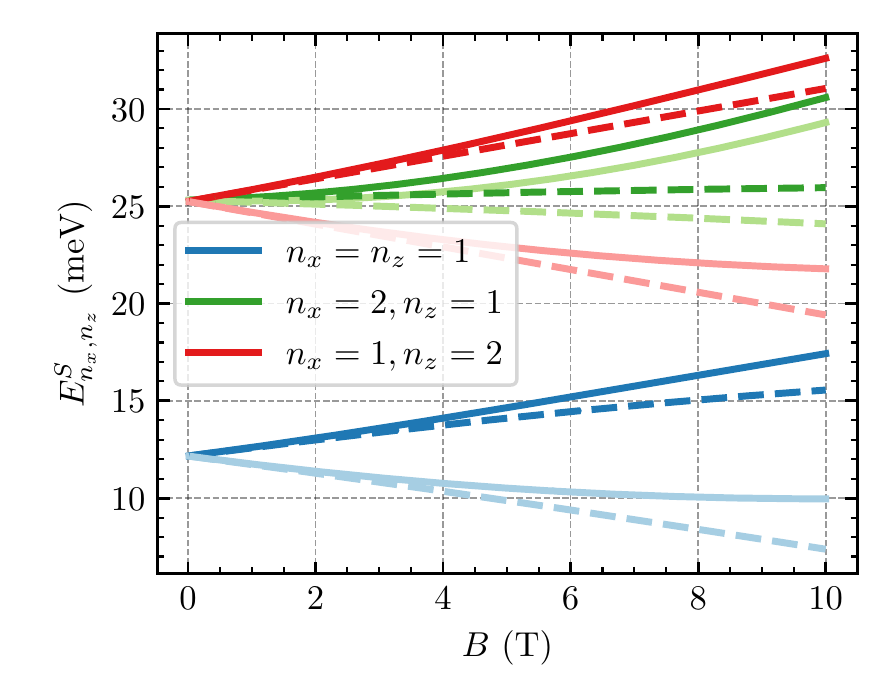}
	\caption{
		The energy levels $E_{n_x,n_z}^{\uparrow,\downarrow}$ of $H_0$, defined in Eq.~\eqref{eq:H0}, of a Ge NW as a function of the perpendicular magnetic field $B$ along the $z$-direction. Here, the parity-mixing term [see Eq.~\eqref{eqn:HintBperp}] is neglected. The dashed lines correspond to Eq.~\eqref{eq:energies_no_orb}, in which we also neglect orbital effects. The solid lines are obtained by the approach developed in Sec.~\ref{sec:orbEff}, where we include orbital effects.  The Zeeman splitting is large ($\approx \SI{1}{\milli\electronvolt}$) at $B = \SI{1}{\tesla}$. The strong renormalization of the $g$-factor due to orbital effects can be observed by comparing the solid and dashed lines. Here, $L_x=L_z=22\,\text{nm}$ and  $E=\SI{1}{\volt\per\micro\meter}$. We also note that  because $L_x=L_z$ the states shown with red and green lines are close in energy at $B=0$. This quasi-degeneracy is a result of the small value of $E$ used here and is lifted in the strong $E$ field limit.
	\label{fig:EVsB_orb_L22}}
\end{figure}

As a starting point for the following perturbation theory, we choose the Hamiltonian
\begin{align}
	H_0=H_{zz}+H_Z+H_E+H_{xy}^{k_y=0}. \label{eq:H0}
\end{align}
Its eigenstates can be approximated in position-space by
\begin{align}
\langle x,z|n_x,n_z,\uparrow \rangle=\left(\begin{matrix}
\sin{(\theta_{n_x,n_z}^{\uparrow}/2)}\phi_{n_z}^\mathrm{H}(z)
\\
0
\\
\cos{(\theta_{n_x,n_z}^{\uparrow}/2)}\phi_{n_z}^\mathrm{L}(z)\\
0
\end{matrix}\right)\phi^{0}_{n_x}(x) \label{eq:Up}
\end{align}
for the $\{3/2,-1/2\}$ subspace, denoted with $S=\uparrow$ in the following,  and by
\begin{align}
\langle x,z|n_x,n_z,\downarrow \rangle=\left(\begin{matrix}
0\\
\cos{(\theta_{n_x,n_z}^{\downarrow}/2)}\phi_{n_z}^\mathrm{L}(z)
\\0
\\
\sin{(\theta_{n_x,n_z}^{\downarrow}/2)}\phi_{n_z}^\mathrm{H}(z)
\end{matrix}\right)\phi^{0}_{n_x}(x) \label{eq:Down}
\end{align}
for the $\{1/2,-3/2\}$ subspace, denoted with $S=\downarrow$.
 Here, we have introduced the angle
 \small
\begin{align}
&\theta_{n_x,n_z}^{\uparrow,\downarrow}=\nonumber\\
&=\arctan\left({\frac{\sqrt{3}\gamma_s\pi^2 n_x^2 \,\langle \phi^\mathrm{H}_{n_z}|\phi^\mathrm{L}_{n_z}  \rangle\,\hbar^2/(L_x^2 m) }{\pi^2 n_x^2 \gamma_s \hbar^2/(L_x^2  m)\pm 4\kappa \mu_B B +\varepsilon^\mathrm{H}(n_z)-\varepsilon^\mathrm{L}(n_z)}}\right).
\end{align}
\normalsize
The corresponding low-energy spectrum $E_{n_x,n_z}^{S}$ is given by
\begin{widetext}
\begin{align}
E_{n_x,n_z}^{\uparrow,\downarrow}=&\frac{\hbar^2\pi^2 n_x^2 \gamma_1}{2 m L_x^2}\pm \kappa\mu_B B+\frac{\varepsilon^\mathrm{H}(n_z)}{2}+\frac{\varepsilon^\mathrm{L}(n_z)}{2}\nonumber\\
&-\frac{1}{2}\sqrt{\left(\frac{\pi^2 n_x^2 \gamma_s \hbar^2}{L_x^2 m}\pm4\kappa \mu_B B +\varepsilon^\mathrm{H}(n_z)-\varepsilon^\mathrm{L}(n_z)\right)^2+\frac{3\gamma_s^2\pi^4n_x^4 |\langle \phi^\mathrm{H}_{n_z}|\phi^\mathrm{L}_{n_z}  \rangle|^2 \hbar^4}{L_x^4 m^2}} \label{eq:energies_no_orb}
\end{align}
\end{widetext}
and is shown in Fig.~\ref{fig:EVsB_orb_L22} for a NW with square cross-section (dashed lines).
Expanding $E_{n_x,n_z}^{\uparrow,\downarrow}$ in the regime of weak electric fields, $L_z/\pi \ll \gamma_1^{1/3} l_E$, we obtain (for square cross-section) an effective $g$-factor along $z$-direction, given by the the following equation
\begin{align}
	&\frac{E_{1,1}^{\uparrow}-E_{1,1}^{\downarrow}}{\mu_B B}=\nonumber\\
	 &\left[4\kappa-\frac{3 L_z^4 m^2 \mu_B^2 B^2 \kappa^3}{\hbar^4 \pi^4\gamma_s^2}+\mathcal{O}\left(B^4\right)\right]\left[1+\mathcal{O}\left(\frac{L_z^6}{l_E^6}\right)\right]. \label{eqn:WOOrbEffLowFieldg}
\end{align} 
From Eq.~\eqref{eqn:WOOrbEffLowFieldg}, we observe that the effective $g$-factor depends on the magnetic field even without accounting for magnetic orbital effects.

The states with $n_x = 2, n_z=1$ and $n_x = 1, n_z=2$ are almost degenerate at $B=0$ in Fig.~\ref{fig:EVsB_orb_L22}. This is due to the fact that  $L_x=L_z$  and that here we use  $E= \SI{1}{\volt\per\micro\meter}$, that is in the weak field limit defined in Sec.~\ref{sec:weakField}. At larger electric fields the quasi-degeneracy is lifted, as expected from Eq.~\eqref{eq:energyZ_E_approx_high2}.

Finally, the parity-mixing term $H_\mathrm{mix} ^{k_y=0}$ couples states with different parity in $x$ (i.e. states with even and odd quantum number $n_x$) and different pseudo-spin $S= \uparrow,\downarrow$.
More explicitly, the  states depicted in blue in Fig.~\ref{fig:EVsB_orb_L22} couple to the  ones in green, and we define the perturbed states
\begin{align}
{|1,1,S \rangle}'=|1,1,S \rangle+\frac{\langle 2,1,\bar{S}|H_\mathrm{mix} ^{k_y=0}|1,1,S\rangle}{E^{S}_{1,1}-E^{\bar{S}}_{2,1}} |2,1,\bar{S} \rangle,
\end{align}
where $\bar{S}$ is the opposite pseudo-spin to $S$, and the overlap 
\begin{align}
	&\langle 2,1,\bar{S}|H_\mathrm{mix} ^{k_y=0}|1,1,S\rangle=\nonumber\\ 
	& \pm - i\frac{8\hbar^3}{3 L_x m}\sqrt{3}\gamma_s \sin\left[{\frac{\theta^{S}_{1,1}+\theta^{\bar{S}}_{2,1}}{2}}\right]  \langle \phi_{1}^\mathrm{H}|k_z|\phi_{1}^\mathrm{L}\rangle.
\end{align} 
In the absence of electric fields, the leading correction to the states $\ket{1,1,S}$ comes from couplings to the states with $n_x = 2$ and  $n_z = 2$.

From the Hamiltonian term linear in $k_y$ [see Eq.~\eqref{eq:linear_no_orb}], we obtain a direct SOI term because the eigenstates defined in Eq.~\eqref{eq:Up} and Eq.~\eqref{eq:Down} (for $n_x=n_z=1$) are directly connected via the terms $\propto k_z k_y$,
\begin{align}
H^{(1)}_{\alpha_{so}}
=i \frac{\hbar^2}{m} \sqrt{3}\gamma_s \sin\left[\frac{\theta^{\uparrow}_{1,1}+\theta^{\downarrow}_{1,1}}{2}\right]\langle \phi^\mathrm{H}_1|k_z|\phi^\mathrm{L}_1 \rangle k_y \sigma_x, \label{eqn:alpha1_noOrb}
\end{align}
where the Pauli matrix $\sigma_x$ acts in the pseudo-spin space $S=\uparrow,\downarrow$.

There is however another sizeable SOI term, that is induced by the linear term $\propto k_x k_y$, and is assisted by the parity-mixing Hamiltonian $H_\mathrm{mix} ^{k_y=0}$. This term is given by
\begin{align}
	&H^{(2)}_{\alpha_{so}} =-i \frac{64 \hbar^6}{3 m^2 L_x^2} \gamma_s^2   
	\langle  \phi^\mathrm{H}_1|\phi^\mathrm{L}_1 \rangle  \langle  \phi^\mathrm{H}_1|k_z|\phi^\mathrm{L}_1 \rangle k_y \sigma_x \nonumber\\
	&  \times \left[ \frac{\sin\frac{\theta^{\uparrow}_{1,1}+\theta^{\downarrow}_{2,1}}{2} \sin\frac{\theta^{\downarrow}_{1,1}-\theta^{\downarrow}_{2,1}}{2}}{E_{1,1}^\uparrow-E_{2,1}^\downarrow} 
	+\frac{\sin\frac{\theta^{\downarrow}_{1,1}+\theta^{\uparrow}_{2,1}}{2} \sin\frac{\theta^{\uparrow}_{1,1}-\theta^{\uparrow}_{2,1}}{2}}{E_{1,1}^\downarrow-E_{2,1}^\uparrow} \right]. \label{eqn:alpha2_noOrb}
\end{align}
The full SOI amplitude is  given by $H_{\alpha_{so}} = H^{(1)}_{\alpha_{so}} + H^{(2)}_{\alpha_{so}} = \alpha_{so} \sigma_x k_y$.

To lowest order in the electric field and while the Zeeman term is small compared to the difference of confinement energies, $ 4\kappa \mu_B B\ll\pi^2 \gamma_s \hbar^2/(L_x^2 m)$, the SOI strength is effectively described by

\begin{align}
	&\alpha_{so}\approx   \frac{\hbar^2}{m}   \frac{L_z^2  \left(\pi^2-6\right) \gamma_s^2}{4 l_E^3 \pi^2 (\gamma_1^2-4\gamma_s^2) r_1}\nonumber\\
	&\times \left[1+
	\frac{128 \gamma_s\left[r_2-4 r_1+2  (r_1-  r_2)L_x^2/L_z^2  \right]}{9 \pi^2 r_2 \left[ 2 \gamma_s (-r_1+r_2)-3 \gamma_1\right]}
	\right], \label{eq:simple_formula}
\end{align}
where we define $r_1=\sqrt{1+L_x^4/L_z^4 - L_x^2/L_z^2}$ and $r_2 = \sqrt{16+L_x^4/L_z^4 - 4 L_x^2/ L_z^2 }$. In particular, in a Ge NW with a square cross-section, we find $\alpha_{so}\approx 0.094 \,e  E L_z^2$. 

\begin{figure*}[tb]
	\includegraphics{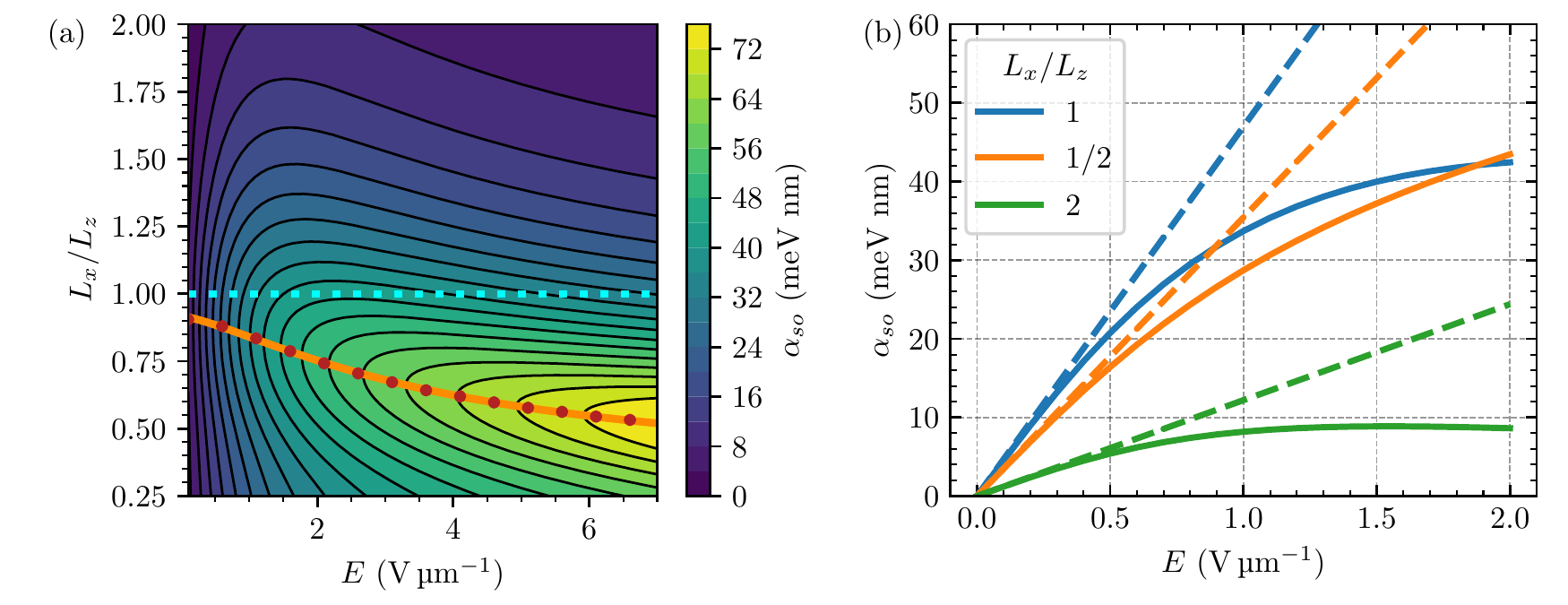}
	\caption{(a) Color code and contour plot of the SOI strength $\alpha_{so}$ obtained from Eqs.~\eqref{eqn:alpha1_noOrb} and~\eqref{eqn:alpha2_noOrb} as a function of perpendicular electric field $E$ and  the confinement length $L_x/L_z$. We fix  $L_z=22\, \text{nm}$ and $B=1\,T$. The red dots mark the maximal SOI and the solid orange line shows the fitting function discussed in the text. 
	With the dotted cyan line we mark the case $L_x=L_z $, where the cross-section is a square.  The plot shows a strong dependence on the geometry of the cross-section, where the SOI increases for decreasing ratio $L_x/L_z$ and increasing $E$-field. (b) Cuts of the plot in (a) for specific side length ratios $L_x/L_z$. The solid lines show the SOI strength according to Eqs.~\eqref{eqn:alpha1_noOrb} and~\eqref{eqn:alpha2_noOrb} and the dashed lines depict the small electric field approximation according to Eq.~\eqref{eq:simple_formula}. The approximation captures very well the linear dependence on $E$ at low field and yields the correct slope for all the considered ratios of side lengths.
	\label{fig:soVsELx_no_orb_L22}}
\end{figure*}
In Fig.~\ref{fig:soVsELx_no_orb_L22}(a), we fix the side length $L_z = \SI{22}{\nano\meter}$ and show a contour plot of the SOI strength $\alpha_{so}$ as a function of the side length $L_x$ and the perpendicular electric field $E$. We see that the square cross-section is not maximizing the SOI strength. In fact, in the figure, the maximal SOI is marked by  red dots and lies on the curve $L_x^m = 2.74 l_E \gamma_1^{1/3} \sqrt{\erf {\left(0.10 L_z^2/(\gamma_1^{2/3} l_E^2)\right)}}$ (solid orange line). This fitting function is similar to the one used in Ref.~\cite{Bosco2021} for the gate-defined Ge 1D channel shown in Fig.~\ref{fig:schematic}(b), where the optimal length is found to be $(L_x^m)_{1D} = 0.81 l_E \gamma_1^{1/3} \sqrt{\erf {\left(0.12 L_z^2/(\gamma_1^{2/3} l_E^2)\right)}}$~\cite{Bosco2021}. Note that for strong electric field ($E>\SI{3}{\volt\per\micro\meter}$) we can neglect the error function and $L_x^m = 2.74 l_E \gamma_1^{1/3}$ has a simple $E^{-1/3}$ dependence. We also find that in the range of parameters considered, our analytical results for the SOI are in good agreement to the more general numerical calculation explained in Sec.~\ref{sec:geff_meff}, also including magnetic orbital effects.

Moreover, in Fig.~\ref{fig:soVsELx_no_orb_L22}(b) we show cuts of the contour plot in Fig.~\ref{fig:soVsELx_no_orb_L22}(a) at certain ratios of the NW side lengths. We compare the result for the SOI strength $\alpha_{so}$ given by Eqs.~\eqref{eqn:alpha1_noOrb} and~\eqref{eqn:alpha2_noOrb} to the low field approximation in Eq.~\eqref{eq:simple_formula} and we observe an excellent agreement at low electric fields. In this case, the SOI strength increases linearly and the approximation yields the correct slope.

\subsection{Solution with orbital effects} \label{sec:orbEff}

In this subsection we account for the orbital effects of the magnetic field. As stated before, we work in the Landau gauge, resulting in $\tilde{\pi}_x = \tilde{k}_x$, $\tilde{\pi}_y = \tilde{k}_y + \tilde{x}$, and $\tilde{\pi}_z =\tilde{k}_z$. To simplify the notation in the following we express the lengths in units of the magnetic length $l_B = \sqrt{\hbar/e B}$, i.e. $\tilde{z}=z/l_B$ and $\tilde{\bf{k}}=l_B \bf{k}$,  and we  introduce the cyclotron energy $\hbar \omega_c=e\hbar B/m$.  Orbital effects renormalize the SOI at large values of $B$, when the side length of the cross-section is comparable to the magnetic length. Here, we discuss a procedure to treat these orbital effects exactly, in contrast to perturbative approaches in other works such as Ref.~\cite{Michal2021}, where a slab geometry is analyzed, or Refs.~\cite{Li2021,Li2021a}, where a cylindrical NW is analyzed. Our results reproduce the effects captured in these works and extend them to the limits examined here. 

Fist, we introduce the ladder operators 
\begin{align}
	a =\frac{\tilde{k}_x - i \tilde{x}}{\sqrt{2}}, \label{eqn:LandauLevelOp}
\end{align}
which satisfy the canonical commutation relation $\left[a, a^\dagger\right] = 1$. 
The Hamiltonian $H_{xy}^{k_y=0}$  [see Eq.~\eqref{eqn:Hxy_ky0}]  can be rewritten as
\begin{widetext}
	\begin{align}
	  	H_{xy}^{k_y=0}= \hbar \omega_c \left(
		\begin{array}{@{}cccc@{}}
			\gamma_+\left(a^\dagger a+\frac{1}{2}\right) & 0 & -\sqrt{3}\gamma_s a^2&0 \\
			0 &\gamma_-\left(a^\dagger a+\frac{1}{2}\right) &  0 & -\sqrt{3}\gamma_s a^2 \\
			-\sqrt{3}\gamma_s (a^\dagger)^2 & 0 & \gamma_-\left(a^\dagger a+\frac{1}{2}\right)&  0  \\
			0 &- \sqrt{3}\gamma_s (a^\dagger)^2 &  0 &\gamma_+\left(a^\dagger a+\frac{1}{2}\right)
		\end{array}
		\right),
	\end{align}
\end{widetext}
\begin{align}
	H_\mathrm{mix} ^{k_y=0}= \sqrt{3} \gamma_s \hbar \omega_c
	\begin{pmatrix}
		0 &  -\sqrt{2} a & 0 & 0\\
		-\sqrt{2} a^{\dagger} & 0 & 0 & 0\\
		0 & 0 & 0 & \sqrt{2} a\\
		0 & 0 & \sqrt{2} a^\dagger & 0
	\end{pmatrix}  \tilde{k}_z. \label{eqn:HintBperp_orbEff}
\end{align}

In analogy with the previous subsection, we now find the eigenstates of $H_0$ [see Eq.~\eqref{eq:H0}]  and then analyze the effect of $H_\mathrm{mix} $ and calculate the SOI strength. In the Landau gauge, the operator 
\begin{align}
	a^\dagger a = \frac{1}{2} \left(-\partial_{\tilde{x}}^2 + \tilde{x}^2 -1\right)
\end{align}
has two eigenfunctions, one even and one odd,  with real-valued eigenvalue $\eta$
\begin{align}
	\label{eq:Eigen-symmetric-LG1}
	\psi^e_\eta(\tilde{x})=\ &e^{-\frac{\tilde{x}^2}{2} } \, _1F_1\left(-\frac{\eta }{2};\frac{1}{2};\tilde{x}^2\right), \\
	\psi^o_\eta(\tilde{x})=\ &i\sqrt{2}\,\tilde{x}\, e^{-\frac{\tilde{x}^2}{2}} \  _1F_1\left(-\frac{\eta}{2}+\frac{1}{2};\frac{3}{2};\tilde{x}^2\right), \label{eq:Eigen-symmetric-LG2}
\end{align}
given in terms of  confluent hypergeometric functions $_1F_1(a, b, \tilde{x})$.
The  annihilation and creation operators act on these eigenfunctions as
\begin{align}
	a\psi^e_\eta&=\eta\psi^o_{\eta-1}, \label{eqn:a_psi_odd}\\
	a\psi^o_\eta&=\psi^e_{\eta-1}, \label{eqn:a_psi_even}\\
	a^\dagger\psi^e_\eta&=(\eta+1)\psi^o_{\eta+1}, \label{eqn:adg_psi_even}\\
	a^\dagger\psi^o_\eta&=\psi^e_{\eta+1}.\label{eqn:adg_psi_odd}
\end{align}
The parity quantum number $\lambda=e,o$ is  a good quantum number at $k_y=0$, when the parity mixing term $H_\mathrm{mix} ^{k_y=0}$ is neglected. In this case, we can express the general solutions for the wave functions in a similar way to Eqs.~\eqref{eq:Up} and~\eqref{eq:Down} with the hypergeometric functions from Eqs.~\eqref{eq:Eigen-symmetric-LG1} and~\eqref{eq:Eigen-symmetric-LG2} substituting the trigonometric functions in the $x$-direction. By using these functions, we solve exactly the Schr{\"o}dinger equation $H_0\Psi=\epsilon\Psi$, with $H_0=H_{zz}+H_Z+H_E+H_{xy}^{k_y=0}$ and with the spinor $\Psi$ being dependent on the general real-valued eigenvalue $\varepsilon$.  The energy of the system is then found by computing the values of $\varepsilon$ for which each component of the wave functions satisfies HW boundary conditions, i.e.  $\Psi_{\uparrow/\downarrow,\mathrm{H/L}}^{\lambda}(L_x/2)=0$.  For more technical details on this analysis and on the  wave functions, we refer to App.~\ref{sec:WF_OE}.

We introduce the new notation $\ket{n_x, n_z, S}$ for these exact solutions. The quantum number $n_x$ labels the possible solutions and determines the parity, and, as before, $S=\uparrow,\downarrow$ is the pseudo-spin.
In Fig.~\ref{fig:EVsB_orb_L22} the solid lines depict the energy levels at $k_y=0$ obtained with this approach. 

In analogy to the analysis in Sec.~\ref{sec:WithoutOrbEff}, the parity-mixing term $H_\mathrm{mix} ^{k_y=0}$ couples states with different pseudo-spin and different parity.
The corrected lowest-energy eigenstates are then
\begin{align}
{|1,1,S \rangle}'=|1,1,S \rangle +\frac{\langle 2,1,\bar{S}|H_\mathrm{mix} ^{k_y=0}|1,1,S\rangle}{E^{S}_{1,1}-E^{\bar{S}}_{2,1}} |2,1,\bar{S} \rangle ,
\end{align} 
and the SOI can be estimated by treating the   terms linear in $k_y$,
\begin{small}
\begin{equation}\label{eq:HLKlinear}
H_{k_y}^\mathrm{orb}=H_{k_y}+\hbar \omega_c \left(\begin{matrix}
\gamma_+ \tilde{x} &0 & \sqrt{3} \gamma_s \tilde{x}  & 0\\
0 & \gamma_-\tilde{x}  & 0 &  \sqrt{3} \gamma_s \tilde{x}   \\
 \sqrt{3} \gamma_s \tilde{x}  & 0 & \gamma_-\tilde{x} & 0 \\
0  &  \sqrt{3} \gamma_s \tilde{x}   & 0  & \gamma_+  \tilde{x}
\end{matrix}\right)\tilde{k}_y,
\end{equation}
\end{small}
perturbatively.
Because the extra terms proportional to $\tilde{x}$ couple states within the same spin subspace states but with different parity in $x$, they contribute to the term assisted by parity mixing $H_{\alpha_{so}}^{(2)}$, see Eqs.~\eqref{eqn:alpha1_noOrb} and~\eqref{eqn:alpha2_noOrb}.
In this case, we find the direct contribution to the SOI 
\begin{align}
	&H^{(1)}_{\alpha_{so}}=\langle 1,1,\downarrow|H_{k_y}|1,1,\uparrow\rangle \sigma_x =
	i \hbar \omega_c \sqrt{3}\gamma_s\nonumber\\
	&\times \left(\langle \Psi_{\downarrow,L}^e|\Psi_{\uparrow,H}^e\rangle+\langle \Psi_{\downarrow,H}^e|\Psi_{\uparrow,L}^e\rangle \right)\langle \phi^{H}_1|\tilde{k}_z|\phi^{L}_1 \rangle \tilde{k}_y \sigma_x \label{eqn:alpha1_Orb}
\end{align}
and the term assisted by parity mixing
\begin{widetext}
\begin{align}
	&H^{(2)}_{\alpha_{so}} =\left(\langle 1,1,\downarrow|H_{k_y}^\mathrm{orb}|2,1,\downarrow \rangle \frac{\langle 2,1, \downarrow|H_\mathrm{mix} ^{k_y=0}|1,1,\uparrow\rangle}{E^{\uparrow}_{1,1}-E^{\downarrow}_{2,1}}+\langle 2,1,\uparrow|H_{k_y}^\mathrm{orb}|1,1,\uparrow \rangle\frac{\langle 1,1, \downarrow |H_\mathrm{mix} ^{k_y=0}|2,1,\uparrow\rangle}{E^{\downarrow}_{1,1}-E^{\uparrow}_{2,1}}\right)\sigma_x\nonumber\\
	&= i  \sqrt{3} (\hbar \omega_c)^2 \gamma_s \bigg\{ \frac{\mel{\phi_1^\mathrm{H}}{\tilde{k}_z}{\phi_1^\mathrm{L}}}{E_{1,1}^\uparrow - E_{2,1}^\downarrow}  \left(\mel{\Psi_{\downarrow, \mathrm{H}}^o}{a^\dagger}{\Psi_{\uparrow, \mathrm{L}}^e} +  \mel{\Psi_{\downarrow, \mathrm{L}}^o}{a^\dagger}{\Psi_{\uparrow, \mathrm{H}}^e}\right)\nonumber\\
	&\times\left[ \gamma_+  \mel{\Psi_{\downarrow, \mathrm{H}}^e}{a-a^\dagger}{\Psi_{\downarrow, \mathrm{H}}^o} + \gamma_- \mel{\Psi_{\downarrow, \mathrm{L}}^e}{a-a^\dagger}{\Psi_{\downarrow, \mathrm{L}}^o} + 2 \sqrt{3} \gamma_s \braket{\phi_1^\mathrm{H}}{\phi_1^\mathrm{L}}\left(\mel{\Psi_{\downarrow, \mathrm{L}}^e}{a}{\Psi_{\downarrow, \mathrm{H}}^o}-\mel{\Psi_{\downarrow, \mathrm{H}}^e}{a^\dagger}{\Psi_{\downarrow, \mathrm{L}}^o}\right) \right]\nonumber\\
	&+\frac{\mel{\phi_1^\mathrm{H}}{\tilde{k}_z}{\phi_1^\mathrm{L}}}{E_{1,1}^\downarrow - E_{2,1}^\uparrow}\left( \mel{\Psi_{\downarrow, \mathrm{H}}^e}{a^\dagger}{\Psi_{\uparrow, \mathrm{L}}^o} + \mel{\Psi_{\downarrow, \mathrm{L}}^e}{a^\dagger}{\Psi_{\uparrow, \mathrm{H}}^o}\right) \nonumber\\
	&\times\left[ \gamma_+ \mel{\Psi_{\uparrow, \mathrm{H}}^o}{a-a^\dagger}{\Psi_{\uparrow, \mathrm{H}}^e} + \gamma_- \mel{\Psi_{\uparrow, \mathrm{L}}^o}{a-a^\dagger}{\Psi_{\uparrow, \mathrm{L}}^e} + 2 \sqrt{3} \gamma_s \braket{\phi_1^\mathrm{H}}{\phi_1^\mathrm{L}} \left(\mel{\Psi_{\uparrow, \mathrm{H}}^o}{a}{\Psi_{\uparrow, \mathrm{L}}^e} -  \mel{\Psi_{\uparrow, \mathrm{L}}^o}{a^\dagger}{\Psi_{\uparrow, \mathrm{H}}^e}\right)\right]\bigg\}\tilde{k}_y \sigma_x . \label{eqn:alpha2_Orb} 
\end{align}
\end{widetext}
The function $\Psi_{\uparrow/\downarrow,\mathrm{H}/\mathrm{L}}^{\lambda}$ is defined in App.~\ref{sec:WF_OE} and $\phi_{n_z}^{\mathrm{H}/\mathrm{L}}$ is given by Eq.~\eqref{eqn:diagBasis_Bperpz_E}. The sum of these two terms yields the total SOI strength $\alpha_{so}$ with effective Rashba-type Hamiltonian, $H_{\alpha_{so}} = H^{(1)}_{\alpha_{so}} + H^{(2)}_{\alpha_{so}} = \alpha_{so} \sigma_x k_y$.

\begin{figure}[htb]
	\includegraphics{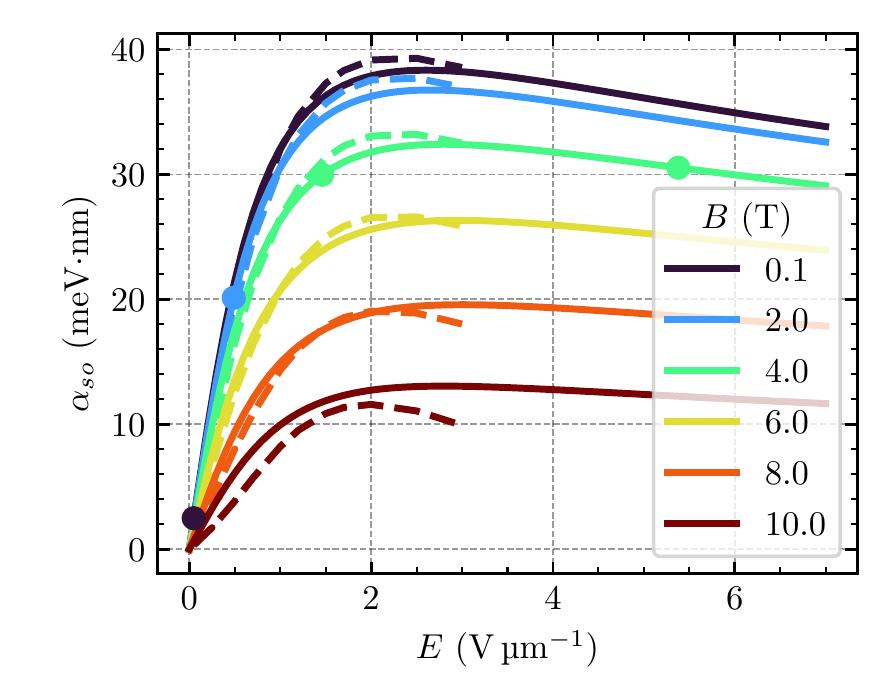}
	\caption{Effective SOI strength $\alpha_{so}$ of a Ge NW with square cross-section, $L_x = L_z =\SI{22}{\nano\meter}$, as a function of the electric field $E$ for different values of the magnetic field. The solid lines are numerical solutions obtained with the discrete basis given in Eq.~\eqref{eqn:basisNumerics}, while the dashed lines show the results from the semi-analytical formulas in Eqs.~\eqref{eqn:alpha1_Orb} and~\eqref{eqn:alpha2_Orb}. For weak electric field the analytical result compares to the numerics very well. The SOI decreases with increasing magnetic field due to orbital effects. The dots mark the points where the ground state dispersion relation becomes flat, and where one needs to include in the effective theory terms that are of higher order in momentum, see Sec.~\ref{sec:QDphys}.\label{fig:SOcouplingOE}}
\end{figure}

As shown in Fig.~\ref{fig:SOcouplingOE}, the SOI strength $\alpha_{so}$ now decreases with magnetic field and the maximum moves towards stronger electric fields. This decrease  can be explained by orbital effects that start to become very relevant at $B\approx\SI{1.4}{\tesla}$, where the magnetic length is comparable to the NW side length. The second term in Eq.~\eqref{eq:HLKlinear} leads to a negative contribution in Eq.~\eqref{eqn:alpha2_Orb} and thus reduces the total SOI $\alpha_{so}$ with increasing magnetic field. In the figure we plot the analytical result (dashed lines) together with numerical results (solid lines). We observe a good agreement between the analytical and numerical curves for weak electric field. The dots in the plot mark where an effective model up to order $k_y^2$ fails because the dispersion relation is dominated by $k_y^4$. This effect will be discussed in detail in Sec.~\ref{sec:QDphys}.

Accounting for anisotropies, $\gamma_1 \neq \gamma_2$, in Ge we see that the SOI only weakly depends on the growth direction of the NW in agreement with Refs.~\cite{Kloeffel2018,Bosco2021}. In particular, the maximum SOI is reached for $z \parallel [110 ],\ \mathrm{NW} \parallel [001]$. This is equivalent to the optimal direct Rashba SOI direction reported in Ref.~\cite{Kloeffel2018}. Comparing this result for the SOI to the result with isotropic approximation for a square NW with side length of $\SI{22}{\nano\meter}$ at $B= \SI{0.1}{\tesla}$ we have a maximum SOI of $\alpha_{so}=\SI{53.8}{\milli\electronvolt \nano\meter}$ at $E= \SI{1.6}{\volt\per\micro\meter}$ [calculated numerically by diagonalizing $H$ in Eq.~\eqref{eqn:FullHamiltonian}] instead of  $\alpha_{so}=\SI{38.3}{\milli\electronvolt \nano\meter}$ at $E= \SI{2.6}{\volt\per\micro\meter}$. Hence, with the right choice of the NW growth direction an even larger SOI at lower electric field than shown by the analytical results before is possible. In contrast to the SOI, the effective $g$-factor strongly depends on the growth direction and is highly sensitive to anisotropies as discussed in detail in Sec.~\ref{sec:beyondSA}.

The calculation of the correction to the $g$-factor coming from the mixing term is inaccurate because it is a second order term in the parity-mixing Hamiltonian [cf. Eq.~\eqref{eqn:HintBperp_orbEff}] and requires to account for many states to converge to the numerical solution. In the following section, therefore, we use a fully numerical approach to calculate the NW quantum levels at $k_y=0$. From this solution, we obtain the effective $g$-factor and then by treating $k_y$ perturbatively, we compute the SOI and the effective masses of the low-energy states.
 
\section{Effective $g$-factor and effective masses \label{sec:geff_meff}}

\begin{figure}[htb]
	\includegraphics[width=\columnwidth]{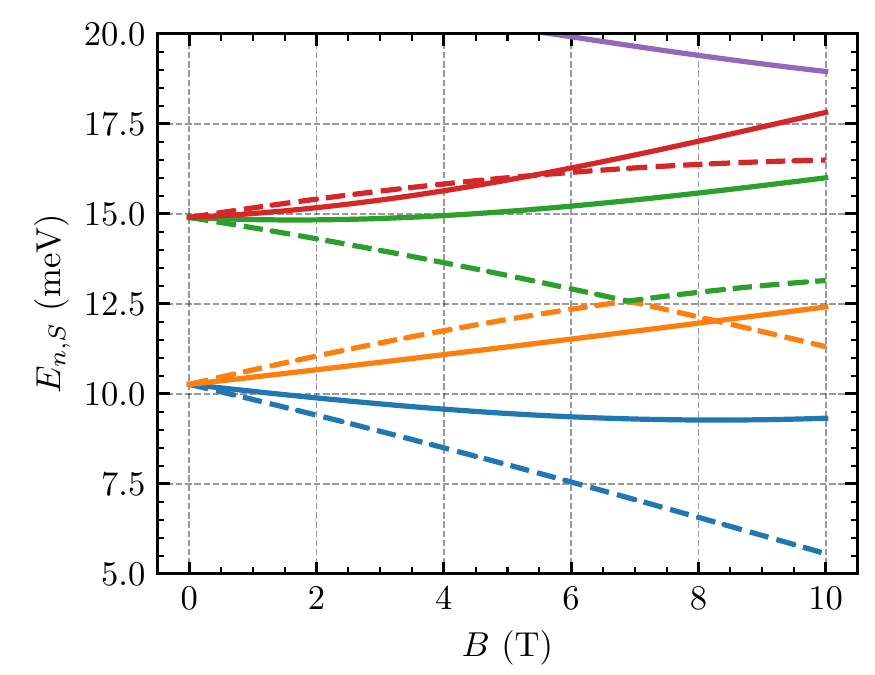}
	\caption{Energy levels of the Hamiltonian in Eq.~\eqref{eqn:FullHamiltonian} calculated numerically by using the discrete basis ($0< n_x, n_z \leq 16$) defined in Eq.~\eqref{eqn:basisNumerics} with (solid lines) and without (dashed lines) orbital effects as a function of the magnetic field. We note that the energy spectrum is substantially modified if orbital effects are taken into account. We consider a square cross-section NW with $L_x = L_z=22\,\text{nm}$, $E=\SI{1}{\volt\per\micro\meter}$, and $\epsilon_s = 0$. Note that in contrast to Fig.~\ref{fig:EVsB_orb_L22} here the parity mixing term defined in Eq.~\eqref{eqn:HintBperp} (without orbital effects) or in Eq.~\eqref{eqn:HintBperp_orbEff} (with orbital effects), respectively, is included. \label{fig:EVsB_L22_numericsE1}}
\end{figure}
In this section we present numerical results for the effective $g$-factor and the effective mass. For the numerical calculations we use the discrete basis 
\begin{align}
	f_{n_x, n_z}(x,z) = \frac{2 \sin\left[n_x \left(\frac{x}{L_x}+ \frac{1}{2}\right)\right] \sin\left[n_z \left(\frac{z}{L_z}+ \frac{1}{2}\right)\right]}{\sqrt{L_x L_z}}  \label{eqn:basisNumerics}
\end{align}
with $0< n_x, n_z \leq 16$. This basis fulfills the HW boundary conditions in $x$- and $z$-directions. We obtain the eigenvalues $E_{n, S}$ of the Hamiltonian in Eq.~\eqref{eqn:FullHamiltonian} with their corresponding eigenfunctions $\psi_{n,S}$, where $n$ labels the states ascending with energy starting from the lowest eigenstate and $S$ is their pseudo-spin. In the following, we extract the parameters of an effective model describing the two states $\psi_{1, \uparrow}(E,B)$ and $\psi_{1, \downarrow}(E,B)$ that are lowest in energy. By second order  perturbation theory, we obtain an effective model Hamiltonian up to order $k_y^2$, 
\begin{align}
	H_\mathrm{eff} = \frac{\hbar^2}{2 \bar{m}}k_y^2 -\beta \sigma_zk_y^2 + g_\mathrm{eff} \frac{\mu_B B}{2} \sigma_z + \alpha_{so} k_y \sigma_x ,\label{eqn:effModel}
\end{align}
with the average effective mass $1/\bar{m}$, the spin-dependent term $\beta$, the effective $g$-factor $g_\mathrm{eff}$, and the SOI strength $\alpha_{so}$. Here, $\beta$ can be interpreted as a spin-dependent mass, which depends on magnetic field and vanishes at $B=0$. Generally, there are further terms possible, such as a diagonal term linear in $k_y$ or an off-diagonal term quadratic in $k_y$. However, these terms are zero in the isotropic LK Hamiltonian~\cite{Hetenyi2020}.
\begin{figure*}[htb]
	\includegraphics{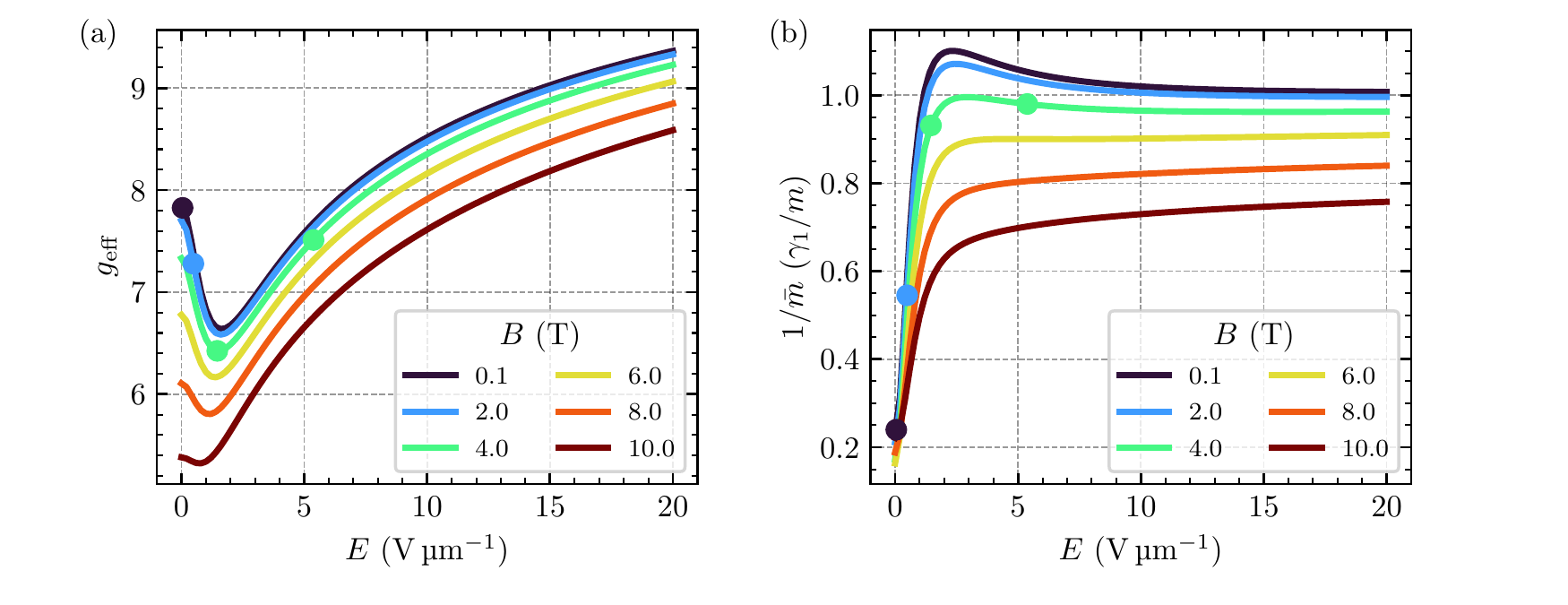}
	\caption{(a) Effective $g$-factor $g_\mathrm{eff}$ from Eq.~\eqref{eqn:geff} and (b) average inverse effective mass $1/\bar{m}$ from Eq.~\eqref{eqn:mbar} of a Ge NW with square cross-section, $L_x = L_z=\SI{22}{\nano\meter}$, as a function of the electric field $E$ at $k_y=0$ for different values of the magnetic field $B$. The $g$-factor is large even at the minimum ($g_\mathrm{eff} > 5$) but it decreases with increasing magnetic field. As described in Sec.~\ref{sec:QDphys}, the minimum is preserved in a QD resulting in a sweet spot where charge noise is suppressed.  The averaged inverse effective mass $1/\bar{m}$ starts from a small value at weak electric fields and approaches a value close to the average HH-LH inverse mass $\gamma_1/m$ at large $E$. The dots mark the points where the ground state dispersion relation becomes flat, and where one needs to include in the effective theory terms that are of higher order in momentum, see Sec.~\ref{sec:QDphys}.\label{fig:effPar_L22}}
\end{figure*}

This effective model works well in different geometries, however in this section we restrict ourselves to the analysis of NWs with square cross-section. We discuss alternative  geometries in Sec.~\ref{sec:Geometries}.
The energy levels at $E=\SI{1}{\volt\per\micro\meter}$ are shown in Fig.~\ref{fig:EVsB_L22_numericsE1} (solid lines). The dashed lines show the same spectrum without orbital effects. Excluding orbital effects the dashed green and orange lines cross, while their solid pendants including orbital effects anti-cross. Thus,  the effective $g$-factor is largely reduced by orbital effects. The large difference between the energies calculated with and without orbital effects leads us to the conclusion that by neglecting orbital effects in Ge NWs one tends to strongly overestimate the $g$-factor. Note also that in contrast to Fig.~\ref{fig:EVsB_orb_L22} here the parity mixing term is fully accounted for. In this case, the parity mixing term leads to a splitting of the at $B=0$ degenerate states with $n_x = 2, n_z=1$ and $n_x = 1, n_z=2$ already in the weak electric field limit. 

More explicitly, we define here the effective $g$-factor for a magnetic field $B$ applied along the $z$-direction as
\begin{align}
	g_\mathrm{eff} = \frac{E_{1, \uparrow}-E_{1, \downarrow}}{\mu_B B}. \label{eqn:geff}
\end{align}
In Fig.~\ref{fig:effPar_L22}(a), we show the dependence of $g_\mathrm{eff}$ in a square-cross-section NW  as a function of the electric field $E$ applied along $z$-direction. The qualitative behaviour is the same at each value of the magnetic field: first, the $g$-factor decreases, it reaches a minimum, and then it grows again. The minimal value of the $g$-factor depends on both electric and magnetic fields and it moves from $g_\mathrm{eff} = 6.6$ at $E= \SI{1.6}{\volt\per\micro\meter}$ for $B=\SI{0.1}{\tesla}$ to $g_\mathrm{eff} = 5.3$ at $E= \SI{0.8}{\volt\per\micro\meter}$ for $B=\SI{10}{\tesla}$. A similar effective $g$-factor of a cylindrical Ge NW for weak $B$ has been predicted in Ref.~\cite{Li2021} and for strong magnetic field in Ref.~\cite{Li2021a}. 

We now analyze the effective mass of the NW, $m_S$. At finite values of the magnetic field, this  parameter depends on the spin $S$ and can be decomposed into the sum of two contributions
\begin{align}
	\frac{1}{m_{S}} = \frac{1}{m_{S}^u} + \frac{1}{m_{S}^p}.
\end{align}
 The first unperturbed ($u$) contribution comes from projecting the part of the LK Hamiltonian, see Eq.~\eqref{eqn:LK_Hamiltonian_sphericalApp}, quadratic in $k_y$,
\begin{align}
	H_{k_y^2} = \frac{\hbar^2}{2m}
	\begin{pmatrix}
		\gamma_+ & 0 & \sqrt{3} \gamma_s & 0 \\
		0 & \gamma_- & 0 & \sqrt{3} \gamma_s \\
		\sqrt{3}\gamma_s & 0 & \gamma_- & 0\\
		0 & \sqrt{3} \gamma_s  & 0& \gamma_+
	\end{pmatrix} k_y^2,
\end{align}
onto the eigenbasis of $H(k_y=0)$ where $H$ is given in Eq.~\eqref{eqn:FullHamiltonian}, 
\begin{align}
	\frac{\hbar^2 k_y^2}{2 m_{S}^u} = \ev{H_{k_y^2}}{\psi_{1, S}}.
\end{align}
The second perturbative ($p$) term is a second-order correction coming from the  term linear in $k_y$,  
\begin{widetext}
\begin{align}
	H_{k_y} = \frac{\hbar^2  \sqrt{3} \gamma_s}{2m}
	\begin{pmatrix}
		\frac{\gamma_+}{ \sqrt{3} \gamma_s} x & i  k_z &(i k_x + x) & 0 \\
		- i k_z & \frac{\gamma_-}{ \sqrt{3} \gamma_s} x & 0 &  (i k_x + x)\\
		(-i k_x + x) & 0 & \frac{\gamma_-}{ \sqrt{3} \gamma_s} x & - i k_z\\
		0 &   (-i k_x + x) & i  k_z & \frac{\gamma_+}{ \sqrt{3} \gamma_s} x
		\end{pmatrix}k_y,
	\end{align} 
\end{widetext}
see e.g. Ref.~\cite{Bosco2021b},
\begin{align}
	\frac{\hbar^2 k_y^2}{2 m_{S}^p} = \sum_{n\geq 2, S'=\uparrow, \downarrow} \frac{\abs{\mel{\psi_{n, S'}}{H_{k_y}}{\psi_{1, S}}}^2}{E_{1, S}- E_{n,S'}}.
\end{align}

At this point, we also define an average effective mass $\bar{m}$ and a spin-dependent term $\beta$ as they appear in the effective model in Eq. ~\eqref{eqn:effModel}:
\begin{align}
	\frac{1}{\bar{m}}&:= {\frac{m_\downarrow + m_\uparrow}{2 m_\downarrow m_\uparrow}} = \frac{1}{2}\left(\frac{1}{m_\uparrow} + \frac{1}{m_\downarrow}\right), \label{eqn:mbar} \\
	\beta &:= -\hbar^2 {\frac{m_\downarrow- m_\uparrow}{m_\downarrow m_\uparrow}} = -\hbar^2\left(\frac{1}{m_\uparrow} - \frac{1}{m_\downarrow}\right). \label{eqn:beta}
\end{align}

The average effective mass $1/\bar{m}$ is shown in Fig.~\ref{fig:effPar_L22}(b) as a function of the electric field. Generally, when  $E$ is large, $1/\bar{m}$ approaches a constant value close to $ \gamma_1/m$, the average HH-LH mass in the LK Hamiltonian. The exact large $E$ limit of $1/\bar{m}$ depends on the magnetic field. Also note that when  below $B<\SI{6}{\tesla}$ $1/\bar{m}$, $1/\bar{m}$ has a maximum, while for $B>\SI{6}{\tesla}$, $1/\bar{m}$ increases monotonically.

\begin{figure}[htb]
	\includegraphics{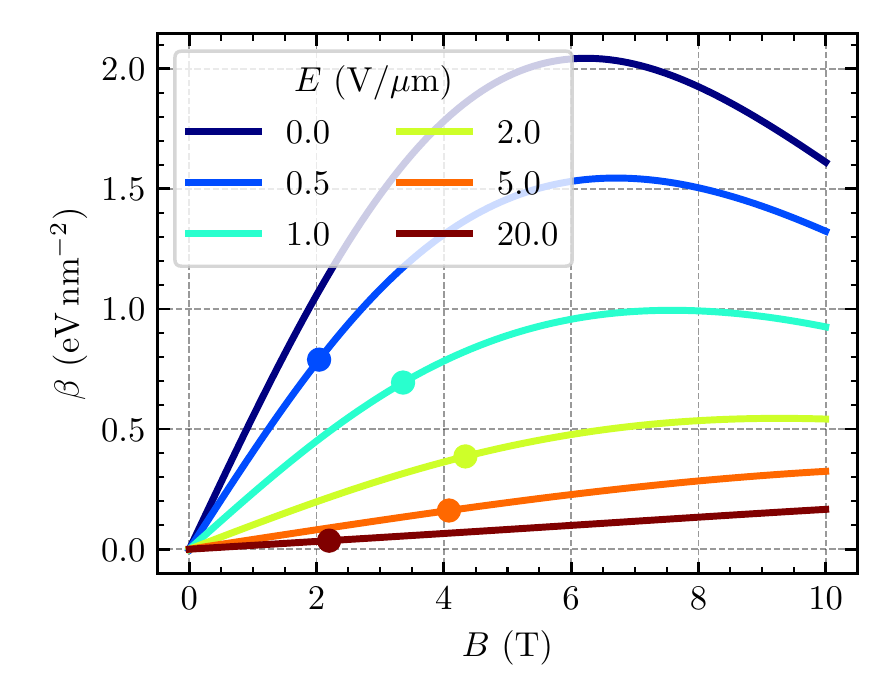}
	\caption{Spin-dependent mass-like term $\beta$ defined in Eq.~\eqref{eqn:beta} as a function of the magnetic field $B$ at $k_y=0$ for different values of the electric field.  At $B=0$, $\beta$ is zero due to time-reversal symmetry and it increases linearly for small $B$. At large $E$, $\beta$ increases very slowly. The dots mark the points where the ground state dispersion relation becomes flat, and where one needs to include in the effective theory terms that are of higher order in momentum, see Sec.~\ref{sec:QDphys}. Here, we use $L_x = L_z= \SI{22}{\nano\meter}$. \label{fig:beta_L22}}
\end{figure}

We show the spin-dependent mass-like term $\beta$ as a function of the magnetic field in Fig.~\ref{fig:beta_L22}. This term is linear in $B$ at low magnetic field and decreases with the electric field. At $B=0$, $\beta=0$ due to time-reversal symmetry. While generally this term is significant, at weak magnetic fields (or at strong electric fields), it  can be justified to consider a simplified effective model with $\beta=0$.

The effective model in Eq.~\eqref{eqn:effModel} is valid when the subband gap is larger than the quantization energy along the $y$-axis. The subband gap including orbital effects is smallest at large values of the magnetic field, see  Fig.~\ref{fig:EVsB_L22_numericsE1}, and at $E=0$ (not plotted). However, in the system considered the gap remains larger than $\SI{1.9}{\milli\electronvolt}$, justifying the use of an effective $2\times 2$-model for sufficiently long QDs.
We note that the subband gap can be increased by reducing the side lengths $L_{x,z}$ or by including strain. Importantly, we also remark that orbital effects extend the validity of this effective model to large values of the magnetic field, and that without them, the effective model can only be valid at weak $B$, far away from the crossing in Fig.~\ref{fig:EVsB_L22_numericsE1}.

\section{Effect of the geometry and confinement details \label{sec:Geometries}}

\begin{figure*}[htb]
	\includegraphics{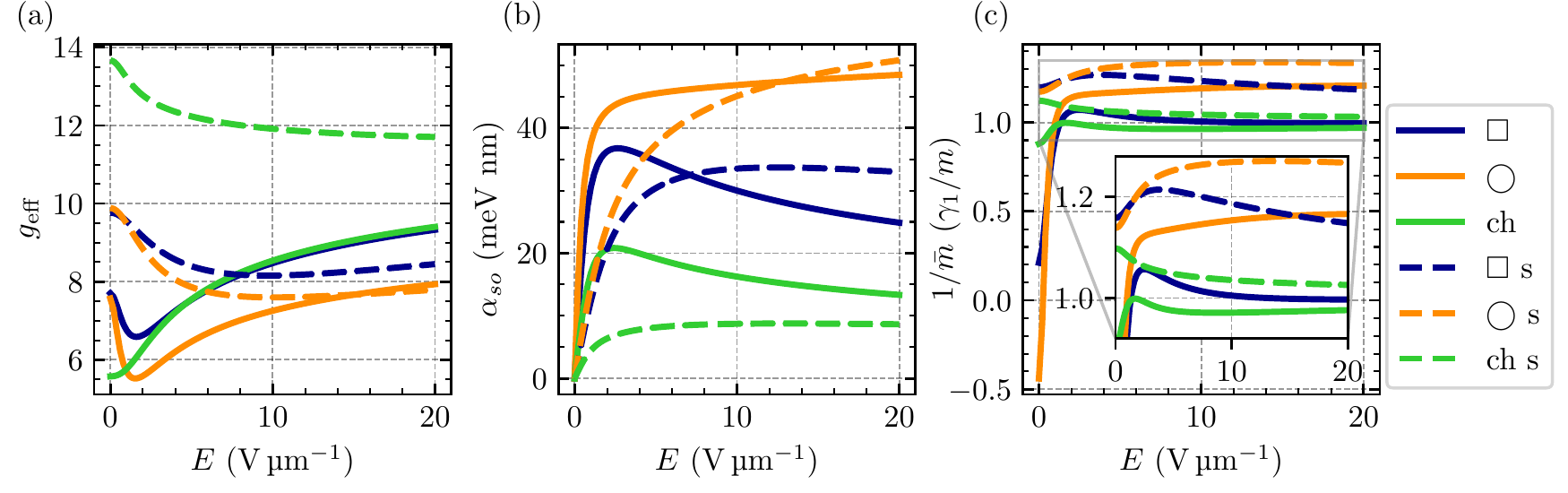}
	\caption{Comparison of the effective parameters of Ge NWs with different geometries with (dashed) and without (solid) strain, obtained by the numerical diagonalization of the Hamiltonian in Eq.~\eqref{eqn:FullHamiltonian}, as described in the text. Here, `$\square$' denotes a NW with a square cross-section (blue) and `$\bigcirc$' a NW with a circular cross-section (orange); `ch' denotes the one-dimensional gate-defined channel (green). With 's' we label strained devices ($\abs{\varepsilon_s} = \SI{0.62}{\percent}$~\cite{Kloeffel2014}, $\varepsilon_s > 0$ for the NW, $\varepsilon_s < 0$ for the channel). (a) The effective $g$-factor $g_{\mathrm{eff}}$. In the NW, this quantity has a minimum that persists even in the presence of strain. (b) SOI strength $\alpha_{so}$. The maximum value of $\alpha_{so}$ is reached at comparably weak electric field without strain. With strain a stronger electric field is required to reach the largest SOI. (c)  The average effective mass $1/\bar{m}$. This quantity tends to converge to a value close to $\gamma_1/m$ and in unstrained NWs with circular cross-section, it is negative at small $E$. Here, $B = \SI{2}{\tesla}$, $L = \SI{22}{\nano\meter}$, $R = L/\sqrt{\pi}\approx\SI{12.4}{\nano\meter}$ and $l_x=L/\pi \approx \SI{7}{\nano\meter}$. \label{fig:effPar_R22_rect-circ}}
\end{figure*}

In the following, we compare the parameters of square NWs with side $L_x=L_z=L$, cylindrical NW with radius $R$ and two-dimensional heterostructures with an electrostatically defined one-dimensional channel, see Fig.~\ref{fig:schematic}(b). In the latter case, the electrostatic potential confining the NW in $x$-direction is
\begin{align}
	U(x) = \frac{\hbar^2\gamma_1}{2 m l_x^4} x^2
\end{align}
and is parameterized by the harmonic  length $l_x$. To describe this case, we use the first 16 eigenstates of the harmonic oscillator. For the NW with circular cross-section with radius $R$, we discretize the cross-section in real space. To compare different cross-sections we choose $l_x = L/\pi$ and $R = L/\sqrt{\pi}$, with $L$ being the side length of the square NW.  

\begin{figure}[htb]
	\includegraphics{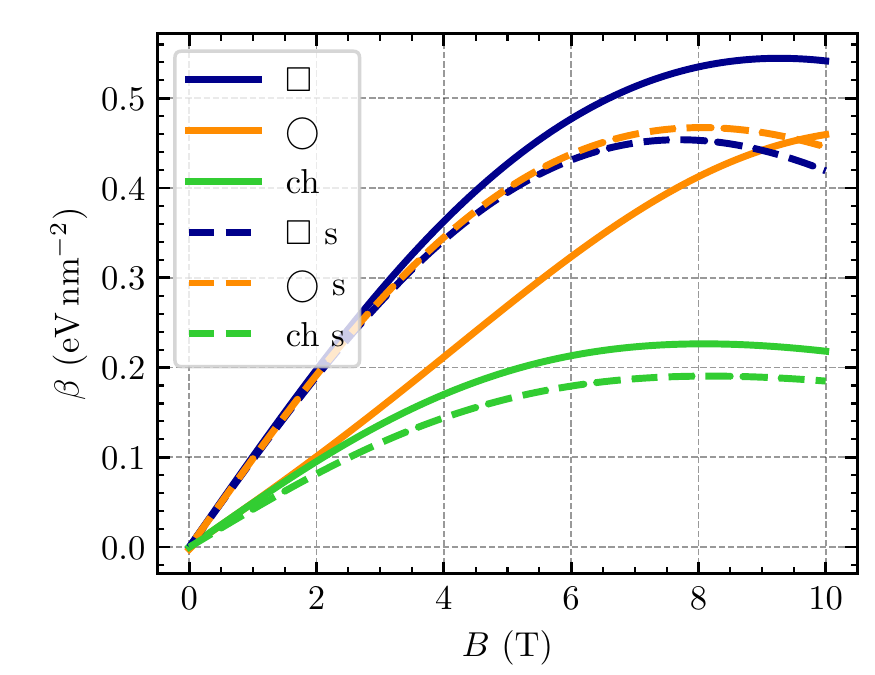}
	\caption{Comparison of the spin-dependent mass-like term $\beta$ in  Ge NWs with different geometries with (dashed) and without (solid) strain as function of perpendicular magnetic field $B$ in $z$-direction. These results are obtained by the numerical diagonalization of the Hamiltonian in Eq.~\eqref{eqn:FullHamiltonian}. Here, `$\square$' denotes the square cross-section, `$\bigcirc$' the circular cross-section, and `ch' the one-dimensional gate-defined channel. With `s' we label the lines with strain ($\abs{\varepsilon_s} = \SI{0.62}{\percent}$~\cite{Kloeffel2014}, $\varepsilon_s > 0$ for the NW, $\varepsilon_s < 0$ for the channel). Time-reversal symmetry at $B=0$ demands $\beta = 0$ . At small $B$ we observe a linear increase regardless of the geometry and strain. Here, $E = \SI{2}{\volt\per\micro\meter}$,  $L = \SI{22}{\nano\meter}$,  $R = L/\sqrt{\pi}\approx\SI{12.4}{\nano\meter}$ and $ l_x=L/\pi \approx \SI{7}{\nano\meter}.$\label{fig:beta_R22_rect-circ}}
\end{figure}

Furthermore, here, we study the effect of strain. For the NW we consider strain induced in the Ge core by a Si shell of relative thickness $\gamma = (L_s-L)/L= 0.1$ ($\gamma = (R_s-R)/R= 0.1$ for cylindrical NW). The strain in the NW is included by the BP Hamiltonian  in Eq.~\eqref{eqn:BPHam}. In contrast, in the two-dimensional heterostructure, the strain is controlled via the percentage of Si in the SiGe layers and it is aligned perpendicularly to the two-dimensional plane as explained in Sec.~\ref{sec:Model}, see in particular Eq.~\eqref{eqn:BPHam_2d}. A comparison between the  effective parameters of the unstrained (solid lines) and strained (dashed lines) devices is shown in Fig.~\ref{fig:effPar_R22_rect-circ}  and Fig.~\ref{fig:beta_R22_rect-circ}. 

The $g$-factors, shown in  Fig.~\ref{fig:effPar_R22_rect-circ}(a), are similar for both NW geometries at weak electric fields and only weakly depend on strain. In this case, the shape of the NW does not play a relevant role because  the wave function is centered in the middle of the cross-section, away from the edges. 
On the one hand, at strong electric fields  the one-dimensional channel resembles a square cross-section NW  because the wave function is compressed at the top of the Ge layer and the parabolic confinement in $x$ becomes less relevant. On the other hand, the difference between circular and square cross-section becomes increasingly important when the wave function is compressed to the top of the NW. 

Moreover, strain generally increases the $g$-factor at weak electric field and moves its minimal value to stronger electric fields. At larger electric fields the $g$-factor of the NWs with strain becomes smaller than without strain. Strain affects much more the one-dimensional channel compared to the NW and it increases the $g$-factor at $E=0$ by almost a factor of three. At strong electric field the values with and without strain are closer to each other, but the strained channel still has a larger $g$-factor. 

In addition, in all cases the SOI strength increases linearly with $E$ at weak electric fields. However, at large $E$, the behaviour of the SOI strength depends on the cross-section, and it either saturates or reaches a maximum before decreasing. These trends are shown in Fig.~\ref{fig:effPar_R22_rect-circ}(b). 
Without strain the SOI for the square NW and the 1D channel reaches a maximum at around $E=\SIrange{2}{ 3}{\volt\per\micro\meter}$, while the cylindrical NW increases monotonically in the whole  range of $E$ studied, from $E=0$ to $E = \SI{20}{\volt\per\micro\meter}$.
In general, strain decreases the SOI at weak electric field since the SOI is inversely proportional to the HH-LH gap, which increases with strain~\cite{Kloeffel2011}. At stronger electric fields, however, the situation changes for the NW devices. In fact, the SOI is larger in the strained NW because the HH-LH gap is decreased by strain (not shown here), and thus, the negative effect of strain on the SOI can be overcome by applying stronger electric fields. The reduction of the SOI due to strain at weak electric field is also reported in Ref.~\cite{Milivojevic2021}.

Next, Fig.~\ref{fig:effPar_R22_rect-circ}(c) shows the average effective mass, $1/\bar{m}$, as a function of the electric field. In analogy to the analysis in Sec.~\ref{sec:geff_meff}, this quantity reaches a value close to $\gamma_1/m$ at strong electric fields, which is slightly increased by strain. At low electric fields, $E<\SI{2}{\volt\per\micro\meter}$, the NWs however present small average masses. In particular, we observe that in  NWs with circular cross-section, the average mass is negative, $1/m_\uparrow<0$. The average mass remains negative at low electric field in a broad range of magnetic fields, from $B=0$ to fields above $B= \SI{10}{\tesla}$. However, even when the mass is negative, there are additional terms that are of higher order in $k$ that ensure the positive curvature of the dispersion relation at large $k$, as discussed in Sec~\ref{sec:beyondHarmApp}.  Thus, even in these cases, it is possible to confine a QD in the NW with an electrostatic potential. Also, in  strained devices, the average mass remains positive. In App.~\ref{sec:SpecNegativeMass}, we highlight the differences in the  dispersion relation of a NW with circular cross-section and having a positive and negative average effective mass.

Finally, in Fig.~\ref{fig:beta_R22_rect-circ} we show the spin-dependent mass term $\beta$. Regardless of the geometry and strain $\beta$ is linear in $B$ at weak magnetic field.  Although this term is typically small, at sufficiently low electric field, it is not negligible and it affects the $g$-factor of an elongated QD created by gating the NW, as shown in the next section. 

\section{Quantum dot physics \label{sec:QDphys}}

\begin{figure*}[htb]
	\includegraphics{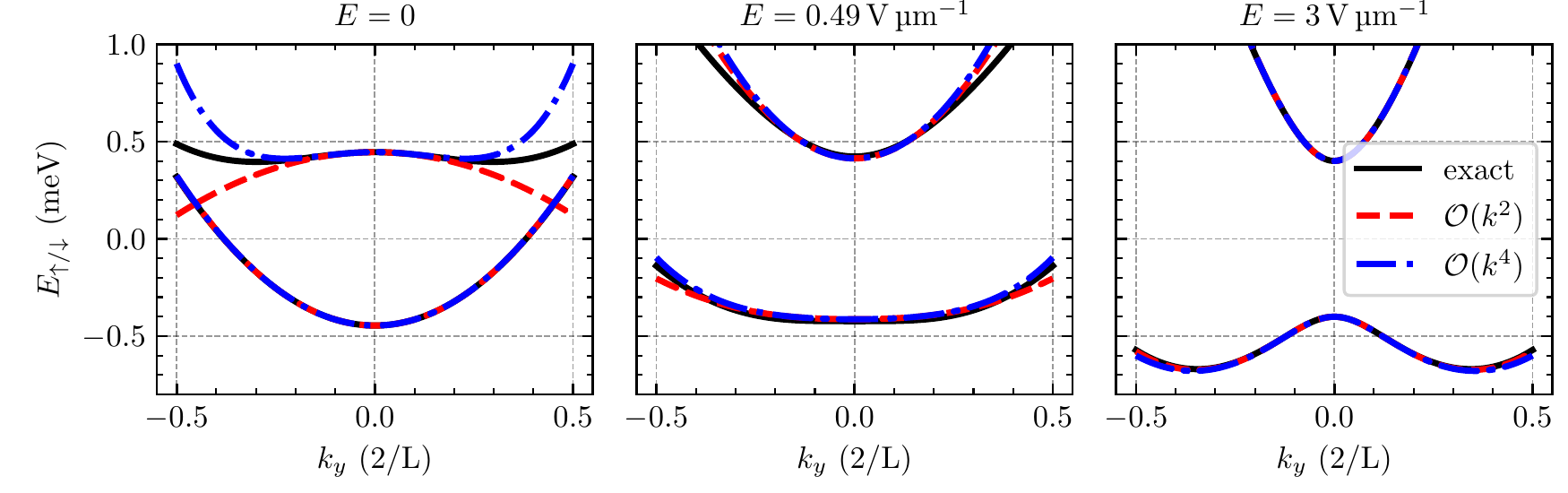}
	\caption{Dispersion relations of a NW with square cross-section of side length $L=\SI{22}{\nano\meter}$ at $B = \SI{2}{\tesla}$ for (a) $E = 0$, (b) $E = \SI{0.49}{\volt\per\micro\meter}$ and (c) $E = \SI{3}{\volt\per\micro\meter}$. The figures compare the exact numerical solution (black) obtained by the diagonalization of the Hamiltonian in Eq.~\eqref{eqn:FullHamiltonian},  the effective model quadratic in $k_y$ (red) in Eq.~\eqref{eqn:effModel}, and the effective model quartic in $k_y$ (blue) discussed in Sec.~\ref{sec:beyondHarmApp}. We observe that the ground state dispersion  is well described by both effective models around $k=0$. Only when the condition from Eq.~\eqref{eqn:flatBandCond} is fulfilled as in (b), the $\mathcal{O}(k_y^2)$ model gives rise to a flat lowest energy band and quartic corrections become more relevant. \label{fig:specK}}
\end{figure*}

This section is dedicated to the physics of a QD in a NW. The dot is defined by an electrostatic confinement potential from gates as sketched in Fig.~\ref{fig:schematic}(a).  First, we focus on the QD $g$-factor and show how strain can be used to tune the position of its minimum such that it occurs at the same electric field where the SOI is maximal. These conditions provide an ideal working point in the parameter space, where the qubit can be driven fast, but at the same time the decoherence rate is diminished by a reduced sensitivity to charge noise~\cite{Vion2002,Petersson2010}. At this sweet spot, we predict ultrafast qubits at low power. In the second part we analyze the effective NW model in Eq.~\eqref{eqn:effModel} for small $k_y$ and show that this model works generally well, except at specific fine-tuned parameters where one needs to include corrections of higher order in momentum. 

In the model given by Eq.~\eqref{eqn:effModel}, the eigenenergies expanded around $k_y=0$ are 
\begin{align}
	E_{\uparrow/\downarrow} = \pm \frac{ E_Z}{2}  + E^{(2)}_{\uparrow/\downarrow} k_y^2 + \mathcal{O}(k_y^4), \label{eqn:GSexpansion}
\end{align}
with Zeeman energy $E_Z=g_{\text{eff}}\mu_B B$, see Eq.~\eqref{eqn:geff}, and
\begin{align}
	E^{(2)}_{\uparrow/\downarrow} &= \frac{\hbar^2}{2\bar{m}}\mp \beta\pm\frac{\alpha_{so}^2}{E_Z}.
\end{align}
In inversion symmetric cross-sections and at zero electric field, the SOI $\alpha_{so}$ is zero~\cite{Bosco2021b}. As the SOI increases with the electric field, the spectrum gradually splits into two separate parabolas with minima at a finite value of $k_y$; at $k_y=0$, these bands are split by $E_Z = g_\mathrm{eff} \mu_B B$. We note that for a certain combination of electric and magnetic fields, the mass of the ground state vanishes, i.e. $E^{(2)}_\downarrow=0$. In particular, this occurs when
\begin{align}
	E^{(2)}_\downarrow = 0 &\Rightarrow E_Z = \alpha_{so}^2 \frac{2 \bar{m}}{\pm \hbar^2+ 2 \beta \bar{m}}\\
	& \Leftrightarrow m_\downarrow = \frac{E_Z \hbar^2}{2 \alpha_{so}^2}. \label{eqn:flatBandCond}
\end{align}

In the vicinity of this point our model quadratic in $k_y$ is not valid and would predict the appearance of a flat band. In this case, we extend our results to fourth order in $k_y$. The points where $E^{(2)}_\downarrow=0$ are marked with dots in Figs.~\ref{fig:SOcouplingOE}, \ref{fig:effPar_L22}, and ~\ref{fig:beta_L22} and the spectrum at one of these points is shown in Fig.~\ref{fig:specK}(b). In this figure, we also show the spectrum for different electric fields, and compare our analytical theory in Eq.~\eqref{eqn:effModel} with numerical results obtained by diagonalizing $H$ in Eq.~\eqref{eqn:FullHamiltonian} using the basis in Eq.~\eqref{eqn:basisNumerics}.
We observe that in general the spectrum is well described by an effective Hamiltonian quadratic in $k_y$. When $E=0$, this effective model fits the ground state dispersion nicely, however, the dispersion of the first excited state is qualitatively correct only up to momenta $\abs{0.3/L}$, and it requires additional higher order corrections for larger $k_y$. Moreover, as anticipated, there are points, e.g. at $E=\SI{0.49}{\volt\per\micro\meter}$, where $E^{(2)}_\downarrow=0$, and the $\mathcal{O}(k_y^2)$ model gives a relatively flat ground state dispersion. In this case, also the exact ground state dispersion is rather flat and can be well described by including terms proportional to $k_y^4$. 

\subsection{Qubit operation}
Having shown the validity of the effective model in Eq.~\eqref{eqn:effModel}, we now use it to analyze a QD. In particular, we study the QD $g$-factor, its sweet spot, and by including an ac electric field $E_y(t)$  applied along the NW, we calculate the frequency of Rabi oscillations induced by EDSR~\cite{Golovach2006}. Here, we consider parameters that are sufficiently far away from  the vanishing effective mass condition in Eq.~\eqref{eqn:flatBandCond}. 
We consider an harmonic confinement potential $V_c = \frac{1}{2}
\bar{m} \omega^2 y^2$ with the harmonic length $l_y = \sqrt{\hbar/(\bar{m} \omega)}$ along the NW. Following Ref.~\cite{Bosco2021} we further introduce the external driving Hamiltonian $H_D(t) = -\hbar k_y \partial_t d_y(t)$ with the time dependent position of the center of the QD $d_y(t)$. We restrict the motion of the QD to the $y$-direction even if the ac field is not perfectly aligned to the NW since the QD is strongly confined in the directions perpendicular to the NW. This leads to the total NW Hamiltonian
\begin{align}
	H_\mathrm{W} = \frac{\hbar^2}{2 \bar{m}} k_y^2 +  \tilde{g} \frac{B}{2} \sigma_z + \alpha_{so} k_y \sigma_x + V_c + H_D(t),
\end{align}
where we introduce $\tilde{g} = \mu_B g_\mathrm{eff} - \beta_0 k_y^2$ with the NW $g$-factor $g_\mathrm{eff}$. Because at weak magnetic fields $\beta \propto B$, we also define the quantity $\beta_0 = \beta/ B$ [cf. Fig.~\ref{fig:beta_L22} and Fig.~\ref{fig:beta_R22_rect-circ}]. The spin-dependent transformation~\cite{Trif2008}
$S = e^{-i \sigma_x y/l_{so}}$ removes the SOI via $S^\dagger (H_\mathrm{eff} +  V_c) S$ where we use the spin-orbit length $l_{so} = \hbar^2/(\bar{m}\alpha_{so})$.

When the confinement energy $\omega$ is much larger than the driving and the Zeeman energy, we obtain an effective QD Hamiltonian by projecting onto the harmonic oscillator ground state at $B = E_y(t) =0$, 
\begin{align}
	H_\mathrm{QD} = \frac{B}{2} \underbrace{e^{-l_y^2/l_{so}^2} \left(\mu_B  g_{\text{eff}} - \frac{\beta_0}{2 l_y^2}\right)}_{=: g_\mathrm{QD} \mu_B}  \sigma_z + \frac{\hbar \partial_t d_y(t)}{l_{so}} \sigma_x. \label{eqn:HamQD}
\end{align}

We now discuss the parameters of the QD theory. 
\begin{figure}[b]
	\includegraphics{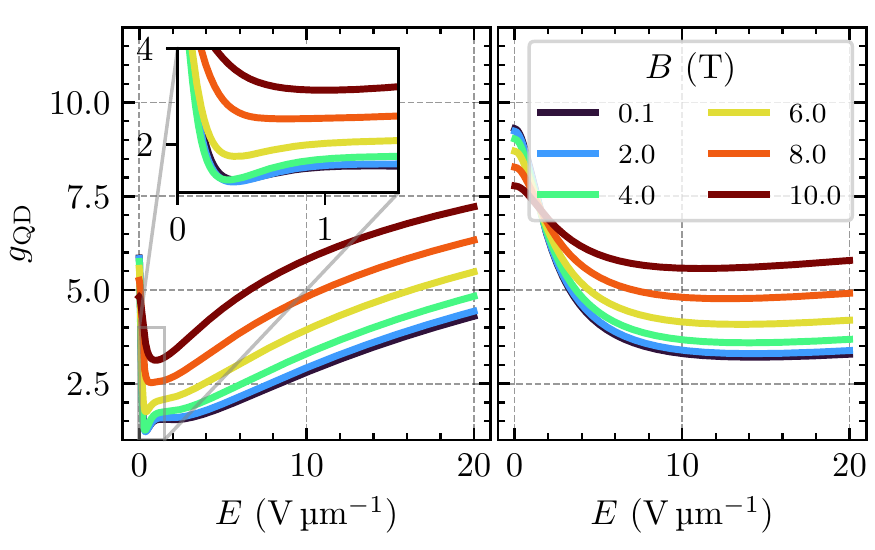}
	\caption{Effective $g$-factor  $g_\mathrm{QD}$ of a Ge NW QD  according to Eq.~\eqref{eqn:HamQD} as a function of the electric field $E$ at $k_z=0$ for different values of the magnetic field $B$. The NW has a square cross-section of  side length $L=\SI{22}{\nano\meter}$ and the QD confinement length is $l_y=\SI{35}{\nano\meter}$. The sweet spot is at very small electric fields in (a), where the NW is unstrained, and it  is shifted to larger electric fields when  strain is included (b). We consider here a strain tensor element $\varepsilon_s = \SI{0.62}{\percent}$. \label{fig:geffQD_L22_l35}}
\end{figure}
In Fig.~\ref{fig:geffQD_L22_l35}, we show the effective renormalized QD $g$-factor $g_\mathrm{QD}$~\cite{Trif2008,Dmytruk2018,Froning2021} for a  QD in a NW with square cross-section of side length $L_x=\SI{22}{\nano\meter}$ and QD confinement length $l_y=\SI{35}{\nano\meter}$. Without strain the QD $g$-factor has a minimum at weak electric field, see Fig.~\ref{fig:geffQD_L22_l35}(a). With strain, as shown in Fig.~\ref{fig:geffQD_L22_l35}(b), the situation is drastically altered. First, in agreement with the analysis in Sec.~\ref{sec:Geometries}, we observe that the QD $g$-factor increases with strain. Furthermore, strain shifts the $g$-factor sweet spot to larger electric fields. We find the minimal value $g_\mathrm{QD} = 3.2$ with strain at the static electric field in $z$-direction $E_\mathrm{sw}= \SI{14.3}{\volt\per\micro\meter}$ instead of  $g_\mathrm{QD} = 1.3$ at $E_\mathrm{sw} = \SI{0.4}{\volt\per\micro\meter}$ and at $B = \SI{0.1}{\tesla}$.

\begin{figure}[htb]
	\includegraphics{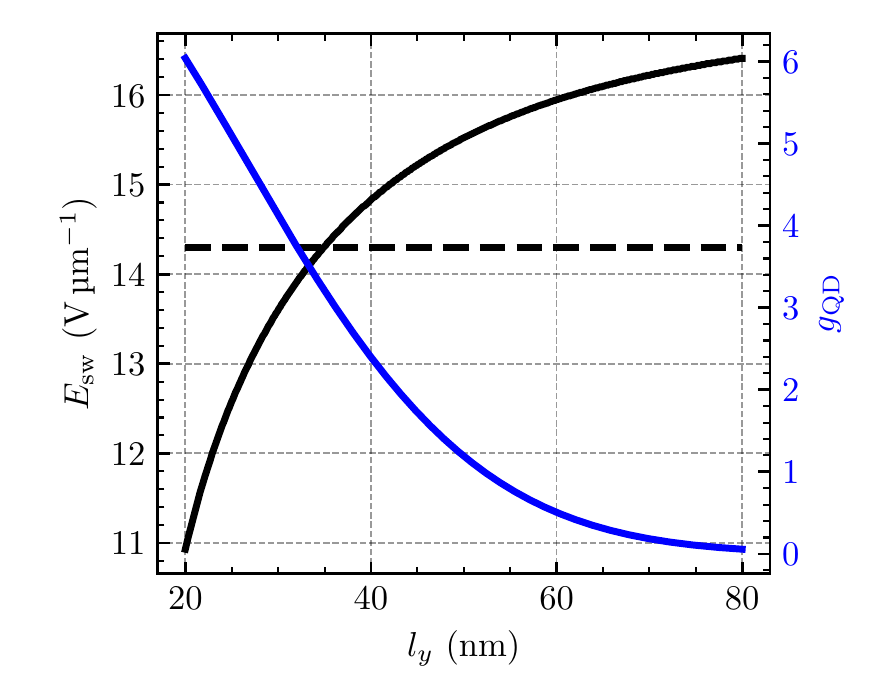}
	\caption{Position of the sweet spot $E_\mathrm{sw}$ (black line) of the effective $g$-factor $g_\mathrm{QD}$ in a strained Ge NW QD as a function of the QD confinement length $l_y$. We consider here $B=\SI{0.1}{\tesla}$ and we calculated $g_\mathrm{QD}$ (see Eq.~\eqref{eqn:HamQD}) numerically by using the discrete basis defined in Eq.~\eqref{eqn:basisNumerics} for diagonalizing $H$ defined in Eq.~\eqref{eqn:FullHamiltonian}. The NW has a square cross-section of  side length $L=\SI{22}{\nano\meter}$. The blue line shows the value of $g_\mathrm{QD}$ at the sweet spot and the horizontal dashed line marks the electric field at which the SOI is maximal. By changing the size of the QD, $l_y$, we can tune the position of the  sweet spot of $g_\mathrm{QD}$ to be at the same value of the electric field at which SOI also achieves its maximum strength. \label{fig:geffbeta}}
\end{figure}

Importantly, including strain the sweet spot of the $g$ factor  occurs exactly at the same value of the electric field that maximizes the SOI ($E= \SI{14.3}{\volt\per\micro\meter}$),  see Fig.~\ref{fig:effPar_R22_rect-circ}(b). Also, we remark that because of the term $\beta_0$ in Eq.~\eqref{eqn:HamQD}, one can tune the position of these sweet spots by  adjusting the size of the QD via the harmonic length $l_y$. The dependence of $g_\mathrm{QD}$ on $l_y$ is examined in detail in Fig.~\ref{fig:geffbeta}, where we show that at certain values of $l_y$, the minimal value of  $g_\mathrm{QD}$ coincides with the optimal SOI.  These working points are ideal for qubit manipulation because they maximize the speed of operation, while reducing drastically the effect of charge noise~\cite{Vion2002,Petersson2010}. However, the qubit is not protected against other sources of noise such as random fluctuations of $l_y$ or hyperfine noise, that could still be addressed by appropriately designing the QD~\cite{Bosco2021a,Bosco2021b}. A detailed analysis of these noise sources in Ge NWs is of interest but is beyond the scope of the present work.

Moreover, when the harmonic drive is given by $E_y (t) = E_{ac} \sin(\omega_D t)$, we can calculate the Rabi frequency $\omega_R$ at the resonance $\omega_D = e^{-l_y^2/l_{so}^2} \left(\mu_B g - \frac{\beta_0}{ l_y^2}\right)   B$ as~\cite{Bosco2021}
\begin{align}
	\omega_R = \frac{l_y}{2 l_{so}}\left(\frac{l_y}{l_E \gamma_1^{1/3}}\right)^3 \frac{E_{ac}}{E}\omega_D.
\end{align}
At the $g$-factor sweet spot and at $B=\SI{0.1}{\tesla}$, we reach with the realistic driving field amplitude $E_{ac} = \SI{0.02}{\volt\per\micro\meter}$ the extremely large Rabi frequency $\omega_R = \SI{2.3}{\giga\hertz}$ with a resonant driving frequency of $\omega_D = \SI{4.5}{\giga\hertz}$. These results indicate that in the setups analyzed here, one can perform ultrafast qubit operations at low power.

\subsection{Beyond the harmonic approximation} \label{sec:beyondHarmApp}
In the following we discuss in more detail the case of $E^{(2)}_\downarrow=0$.
Including the term $ H_4 = \left(\frac{A_+}{2} +\frac{A_-}{2}\sigma_z\right) k_y^4 $, coming from fourth order perturbation theory in $k_y$, the eigenenergies modify as
\begin{align}
E_{\uparrow/\downarrow} = \pm \frac{ E_Z}{2} + \left(\hbar^2 A_\downarrow + \frac{\hbar^2 \alpha_{so}^2}{2 E_Z^2 \abs{m_\uparrow}}\right)  k_y^4 + \mathcal{O}(k_y^6), \label{eqn:solk4}
\end{align}
where we  define $A_{\downarrow/\uparrow} = (A_+ \mp A_-)/2$. The Hamiltonian $H_4$ completely determines the spectrum when $E^{(2)}_\downarrow=0$. The spectrum given by the effective model in Eq.~\eqref{eqn:effModel} including the term $H_4$ is shown in Fig.~\ref{fig:specK} with blue lines. The values of the parameter $A_\uparrow$ and $A_\downarrow$ used in the figure are given in Tab.~\ref{tab:Avals}. For large $k_y$, the quartic Hamiltonian gives a better estimate for the ground state dispersion compared to quadratic model in Eq.~\eqref{eqn:effModel}.
\begin{table}[t]
	\caption{Explicit values for $A_\uparrow$ and $A_\downarrow$ used in Fig.~\ref{fig:specK} calculated numerically in fourth order perturbation theory. \label{tab:Avals}}
	\begin{ruledtabular}
		\begin{tabular}{c|ccc}
			& $E=0$ & $E=\SI{0.49}{\volt\per\micro\meter}$ & $E=\SI{3}{\volt\per\micro\meter}$ \\\hline\hline
			$A_\uparrow$ $[\si{\electronvolt\nm^4}]$ & $0.358$ & $-8.19$ & $-4.61$ \\\hline
			$A_\downarrow$ $[\si{\electronvolt\nm^4}]$& $183$ & $55.3$ & $-5.88$ 
		\end{tabular}
	\end{ruledtabular}
\end{table}

We point out that when the mass vanishes, there are many energy states that are close in energy, and, thus we envision that this regime could be interesting for simulations of strongly correlated matter, e.g.  the SYK model~\cite{Sachdev1993,Sachdev2015,Altland2019}.
 We now estimate the number of states we can put into such a QD. To do so, we consider an harmonic QD confinement $V_c y^2 = \frac{1}{2} \bar{m}  \omega^2 y^2$ with the confinement length $l_y = \sqrt{\hbar/(\bar{m}\omega)}$ along the NW. Then we solve the differential equation
\begin{align}
	(E^{(2)}_\downarrow k_y^2 + E^{(4)}_\downarrow k_y^4 + V_c y^2) \phi(k_y) = E\phi( k_y)
\end{align}
numerically and find its lowest eigenvalues. This procedure allows us to see how many states can coexist in the QD below a certain energy threshold $E_T= k_B T$ set by the temperature. In what follows we  focus on the number of states at $T=\SI{1}{\kelvin}$. With a confinement length of $l_y = \SI{186}{\nano\meter}$~\footnote{We obtain this value from explicitly solving the Laplace equation for a three-gate-setup with gate side length $\SI{200}{nm}$ , distance between the gates $\SI{100}{nm}$, distance between the QD and the gates $\SI{120}{nm}$, center gate voltage $\SI{50}{\milli\volt}$ and side gate voltages $\SI{55}{\milli\volt}$} we obtain $V_c = \SI{4.25E-7}{\milli\electronvolt\per\square\nano\meter}$; we consider also a NW with side length $L=\SI{15}{\nano\meter}$ and magnetic field  $B=\SI{2}{\tesla}$, such that  the condition for vanishing mass in Eq.~\eqref{eqn:flatBandCond} is fulfilled at $E = \SI{0.72}{\volt\per\micro\meter}$. With these parameters we have in total eleven single-hole states in the QD in contrast to a QD with typical effective mass $m/\gamma_1$, where under the same conditions, one obtains three states below $E_T$. A detailed analysis of hole-hole interactions in these system is an interesting problem for future work, but it goes beyond the scope of the present paper.

\section{Beyond the isotropic approximation \label{sec:beyondSA}} 

In this Section, we analyze the limits of the isotropic approximation used in the previous sections and commonly adopted in literature~\cite{Li2021,Li2021a,Michal2021,Milivojevic2021}. As we show in Sec.~\ref{sec:geff_meff}, at low magnetic fields only the effective $g$-factor is significantly affected by orbital effects. Thus, we focus on the effective NW $g$-factor and how it depends on the growth direction with and without orbital effects. 

We use here the general anisotropic LK Hamiltonian 
\begin{align}
	H_\text{LK} = &\frac{\hbar^2}{2 m} \left[\gamma_k \vect{\pi}^2 -2 \gamma_2 (\pi_{x'}^2J_{x'}^2 + \pi_{y'}^2 J_{y'}^2 +\pi_{z'}^2 J_{z'}^2)\right.\nonumber\\
	&\left. \vphantom{\pi^2_x}\! - 4 \gamma_3 \left(\{\pi_{x'}, \pi_{y'}\}\{J_{x'}, J_{y'}\} + \text{c.p.}\right)\right] \label{eqn:LK_Hamiltonian}
\end{align}
where the primed indices $x', y', z'$ denote the axes aligned to the main crystallographic axes $[100]$, $[010]$, and $[001]$, respectively. Strain is included via the isotropic BP Hamiltonian introduced in Eq.~\eqref{eqn:BPHam}, and neglecting corrections coming from different growth directions~\cite{Kloeffel2014}.
We consider the coordinate system specified in Fig.~\ref{fig:schematic} including orbital effects via the Landau gauge $\vect{A} = (0, x, 0) B$. Since we are mainly interested in the orbital effects of the magnetic field, we consider at first $E=0$. To account for different growth directions we rotate the LK Hamiltonian and solve for the eigenvalues numerically as described in Sec.~\ref{sec:geff_meff}. For further information on the rotations see App.~\ref{sec:rotations}.

\begin{figure*}[htb]
    \includegraphics{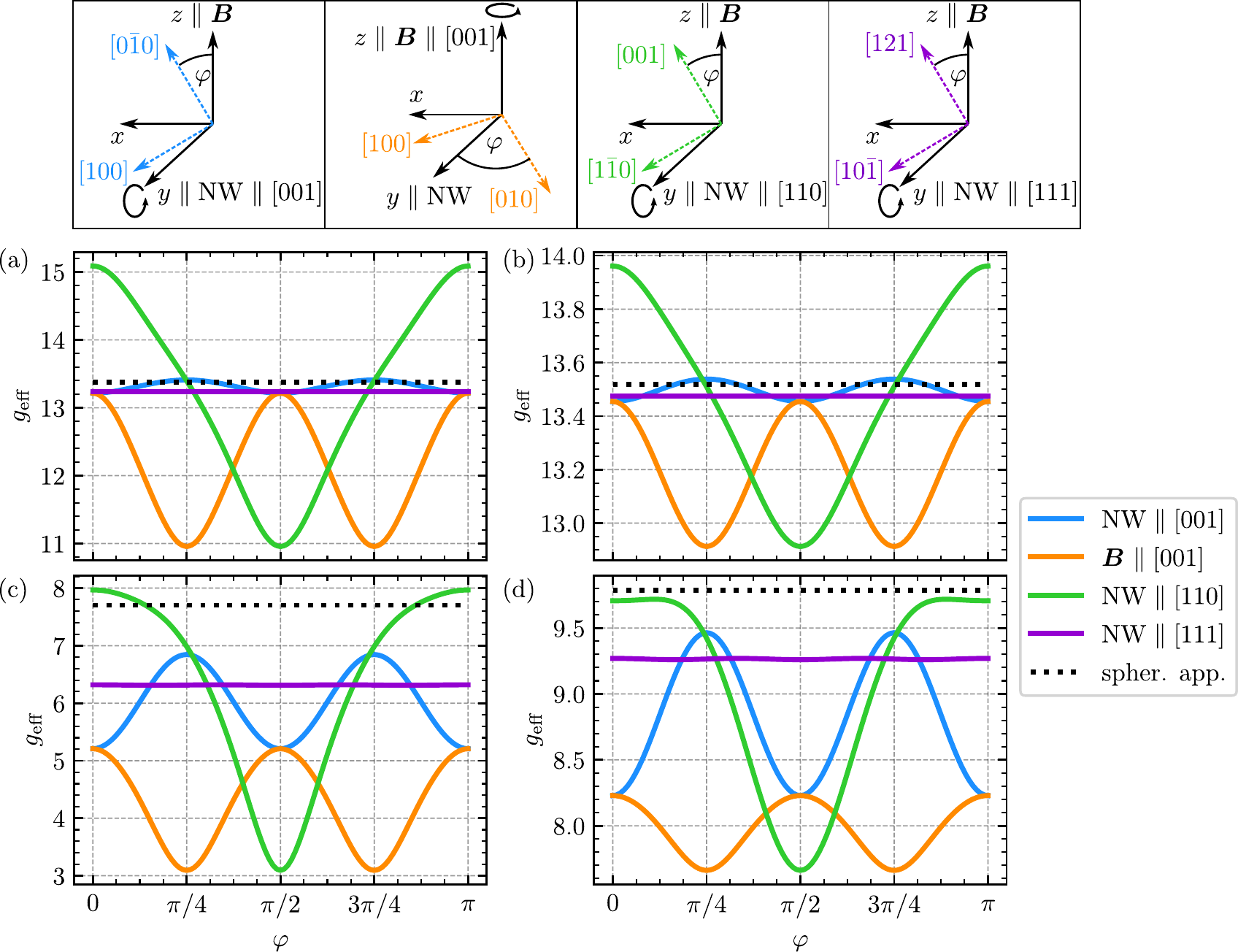}
	\caption{Effective $g$-factor $g_{\mathrm{eff}}$ for different growth directions of the NW, obtained numerically from the diagonalization of the Hamiltonian given by  Eq.~\eqref{eqn:FullHamiltonian}  and the anisotropic LK Hamiltonian given by Eq.~\eqref{eqn:LK_Hamiltonian}, using the discrete basis defined in Eq.~\eqref{eqn:basisNumerics}. The top figures show the rotations of the coordinate system corresponding to the color code in the legend. In each case one axis is fixed and then we rotate by the angle $\varphi$ around it. For the blue, green and purple lines the NW-axis ($y$) is fixed parallel to the crystallographic directions $[001]$, $[100]$, and $[111]$, respectively. For the orange line, we fix $z \parallel [001]$. (a) Without orbital effects, without strain; (b) without orbital effects, with strain; (c) with orbital effects, without strain; (d) with orbital effects, with strain. The $g$-factor depends on the growth direction and this dependence changes when orbital effects are included. This leads to the conclusion that the isotropic approximation is not well justified in Ge NWs in a general case without strain. In core/shell NWs with strain the isotropic approximation is better justified. We choose a NW with a square cross-section with side length $L_x = L_z = \SI{22}{\nano\meter}$, $B=\SI{0.1}{\tesla}$, $E=0$, and strain tensor element $\varepsilon_{s} = \SI{0.62}{\percent}$.  \label{fig:growthDir}}
\end{figure*}

In Fig.~\ref{fig:growthDir}, we show how the effective NW $g$-factor $g_{\text{eff}}$ depends on the growth direction of the NW. As in the previous sections, we consider a NW  parallel to $y$ and $\vect{B}\parallel z$, see Fig.~\ref{fig:schematic}. In particular, we analyze the four cases shown at the top of Fig.~\ref{fig:growthDir}: first, we consider three typical growth directions of a NW ($y\parallel [001]$, $[110]$, and $[111]$) and, second, we consider a magnetic field  aligned to a main crystallographic axis, i.e. ($B \parallel  [001]\parallel z$). 
We also rotate the coordinate system by an angle $\varphi$ according to Tab.~\ref{tab:growthDirections}. More specifically, by varying $\varphi$ in the first case the magnetic field points along different crystallographic axes, and in the second case the growth direction of the NW changes [Note that to improve readability, in Tab.~\ref{tab:growthDirections}, Tab.~\ref{tab:growthDirectionsExplicit}, and App.~\ref{sec:rotations}, we choose a different coordinate system compared to here. In particular, the coordinate system in the main text is related to the one in the appendix by the replacements $x \rightarrow y,\ y\rightarrow z,\ z \rightarrow x$].

The orange curve in Fig.~\ref{fig:growthDir} shows the $g$-factor when the magnetic field $\vect{B}\parallel z$ is aligned to one of the main crystallographic axes, i.e. $\vect{B} \parallel  [100],[010],[001]$, while the blue curve shows $g_{\text{eff}}$ when the NW is aligned to these axes, i.e. $y \parallel  [100],[010],[001]$. At the angles  $\varphi=0, \pi/2, \pi$, these two curves describe the same situation, and thus the values of $g_{\text{eff}}$ coincide. In analogy, the orange curve at $\varphi=\pi/4,3\pi/4$ describes a NW grown along the $[110]$-direction with a magnetic field aligned to a main crystallographic axis; this case is equivalently described by the green line at $\varphi=\pi/2$.

Moreover, the dotted black curve shows the values of the NW $g$-factors in the isotropic approximation, and is independent of the angle $\varphi$. In the other cases, however, we expect the $g$-factor to be an oscillating function of $\varphi$. These oscillations can have a rather large amplitude when $y\parallel [001]$ and $y\parallel [110]$. Without orbital effects, see Fig.~\ref{fig:growthDir}(a), $g_{\text{eff}}$ varies  at most of $\pm \SI{20}{\percent}$ from the $g$-factor obtained with the isotropic approximation, and therefore,  this approximation is justified in this case.

To have a better understanding of the origin of these oscillations, we refer to App.~\ref{sec:rotations}, where we report explicit expressions of the LK Hamiltonian in all the cases considered here [note the different coordinate system used in the Appendix]. 
In particular, when $\mathrm{NW}\parallel[001]$, the LK Hamiltonian in Eq.~\eqref{eqn:LKHam_z001} has a term  $\propto e^{\pm 4 i \varphi}$ that modulates the amplitude of the coupling between spin $\pm 3/2$ to spin $\mp 1/2$.
This modulation leads to a change of the HH-LH mixing with $\varphi$ and, thus, to an oscillation of the $g$-factor with a periodicity of $\pi/2$. The maximum LH contribution to the ground state is obtained at $\varphi=\pi/4$ where the $g$-factor is minimal. 
In analogy, the orange curves corresponding to the case where $\vect{B} \parallel z \parallel[001]$ have a similar $\varphi$-dependence, which can also be explained  by a term $\propto \cos(4 \varphi)$. In contrast, when $\mathrm{NW}\parallel [110]$ (green curve) the $\varphi$-dependent coupling between spins $\pm 3/2$ and $\mp 1/2$ is $\propto e^{\pm i \varphi}$, see Eq.~\eqref{eqn:LKHam_z110}, resulting in a $\pi$-periodic $g_{\text{eff}}$.

Interestingly, we note that the oscillations disappear when $y\parallel [111]$ (purple line).
In this case, in fact, the rotation angle $\varphi$  only changes the phase of the HH-LH matrix elements in the LK Hamiltonian, but it does not affect  their amplitude, see  Eq.~\eqref{eqn:LKHam_z111}. Consequently $\varphi$ does not modify the $g$-factor at $k_y=0$. 

We obtain a similar result with strain [cf. Fig.~\ref{fig:growthDir}(b)]. The deviation from the $g$-factor with isotropic approximation (dashed lines in Fig.~\ref{fig:growthDir}) is less than $\SI{5}{\percent}$. Hence, with strain the isotropic approximation is even more justified than in the unstrained case. This difference can be explained by the larger subband gap between the two lowest energy states and the excited states with strain~\cite{Kloeffel2018}. Interestingly, with and without strain, the $g$-factor for $\mathrm{NW}\parallel [110]$ is closest to the isotropic approximation for $\varphi = \pi/4, 3\pi/4$ corresponding to $\vect{B}$ being parallel to highly non-symmetric directions.

Including orbital effects into our calculations changes drastically the picture. Without strain [cf. Fig.~\ref{fig:growthDir}(c)], for specific angles the $g$-factor can vary up to $\SI{60}{\percent}$ from the isotropic approximation. This large variation occurs when $\vect{B}\parallel [001]$ with $\varphi = \pi/4 $ and $\varphi = 3\pi/4 $, where $x\parallel [110]$, $\mathrm{NW}\parallel [\bar{1}10]$, and  $x\parallel [1\bar{1}0]$, $\mathrm{NW}\parallel [110]$, respectively, and $\mathrm{NW}\parallel [110]$ with $\varphi=0$ where $\vect{B}\parallel [1\bar{1}0]$, $x\parallel [00\bar{1}]$, and $\varphi=\pi$ where $\vect{B}\parallel [\bar{1}10]$, $x\parallel [001]$. 
With such a large deviation the isotropic approximation is not well justified anymore to describe the $g$-factor correctly. For $\mathrm{NW}\parallel [111]$ the $g$-factor oscillates now slightly with the angle between $6.31$ and $6.32$ due to an additional angle dependence in the orbital effect terms that couple spin $\pm 3/2$ to spin $\mp 1/2$ states (see Eq.~\eqref{eqn:LKHam_z111}). Because of the triangular symmetry of the crystal when $\mathrm{NW}\parallel [111]$, the periodicity is $\pi/3$.

Additionally including strain into our calculations generally increases the $g$-factor and brings it closer to the $g$-factor with isotropic approximation. In this case, the maximum deviation from the $g$-factor with isotropic approximation is around $\SI{20}{\percent}$. These results show the profound impact of the magnetic orbital effects even at weak magnetic fields, and they show how big the influence of strain can be. 

In conclusion our results show that without strain the isotropic approximation is justified only in special cases where the effective $g$-factor depends only weakly on the rotation angle. With strain the results for the $g$-factor deviate less form the result with isotropic approximation, and thus increasing the strain by using a thicker Si shell would render the Ge core even more isotropic.

\section{Conclusion} \label{sec:conclusion}

We  derived a low-energy effective one-dimensional model that describes the physics of confined hole systems when electric and magnetic fields are applied in one of the confined directions.
We developed an analytic approach to quantify the spin-orbit interactions of this model, and, assisted by numerical calculations, we investigate the dependence of SOI, $g$-factor, and effective masses on the applied fields, on strain and on the magnetic orbital effects. These effects are crucial to have a good description of the system.
 
 In particular, by complementing the analytical approach with numerical calculations we find that the $g$-factor is strongly renormalized by orbital effects even at low magnetic fields. Moreover, the orbital effects introduce a strong dependence of the $g$-factor on the material growth direction, which leads to a break-down of the isotropic approximation typically employed for Ge.

We find an excellent agreement between analytically and numerically computed SOI in the weak electric field limit. At strong fields, our analytical theory captures the qualitative trend of the numerical results, but is quantitatively imprecise. We also find that the SOI decreases with increasing magnetic field. Our analysis enables a better understanding of approximately isotropic semiconductor NWs including orbital effects. 
Moreover, we show that a square cross-section is not the best choice for optimizing the SOI and that the optimal NW cross-section is rectangular with a width that depends on the electric field as $L_x \approx 2.74 l_E \gamma_1$. We also identify an extra term that can be interpreted as a spin-dependent effective mass.

The analysis of different NW geometries reveals that at low electric field circular and square cross-section are very similar, while in the strong field limit (typically reached for $E>\SI{3}{\volt\per\micro\meter}$) a gate defined one-dimensional channel is comparable to the square cross-section. Furthermore, we analyze the influence of strain and observe that it increases the $g$-factor and reduces the SOI at weak electric field. 

We show that in a QD in qubit operation mode it is possible to tune the SOI maximum and the $g$-factor sweet spot to be at the same electric field by designing strain and confinement potential. At the sweet spot we predict Rabi frequencies in the \si{\giga\hertz} range at low power, enabling  ultrafast gates. With this result it is possible to optimize electrically controlled qubits in Ge NW QDs and we believe a similar optimization is possible in other approximately isotropic semiconductor NWs.

The effective model, Eq.~\eqref{eqn:effModel}, we present is valid for most of the relevant NW growth directions. 
We discuss the growth direction dependence of the $g$-factor and show that orbital effects play an important role, even at low magnetic field, but that they can be counteracted by strain.
Finally, we observe that the effective NW model can break down at certain electric and magnetic fields, resulting in a flat band over $k$. In these cases the physics is dominated by a $k^4$ term. This interesting working point could open up the possibility of investigating strongly correlated systems in QDs.

\begin{acknowledgments}
We would like to thank  Christoph Kloeffel for useful
discussions and comments. This work is supported by the
Swiss National Science Foundation (SNSF) and NCCR SPIN.
M.B. acknowledges support by the Georg H. Endress Foundation.
This project received funding from the European Union's Horizon 2020 research and innovation program (ERC Starting Grant, grant agreement No 757725).
\end{acknowledgments}

\appendix

\section{Wave functions with orbital effects \label{sec:WF_OE}}
In this Appendix we provide some details on the wave functions in which orbital effects are included exactly.  If we neglect the parity mixing term $H_\mathrm{mix} ^{k_y = 0}$, the general solutions for the wavefunctions of $H_0=H_{zz}+H_Z+H_E+H_{xy}^{k_y=0}$ with orbital effects are written as
\begin{align}
	\Psi_\uparrow^\lambda (x, z) &=\left(
	\begin{array}{c}
		\phi^\mathrm{H}_{n_z}(z) \Psi_{\uparrow, \mathrm{H}}^\lambda(x)\\
		0\\
		\phi^\mathrm{L}_{n_z}(z) \Psi_{\uparrow, \mathrm{L}}^\lambda(x)\\
		0
	\end{array}
	\right), \label{eqn:diagBasis_Bperp11}\\
	\Psi_\downarrow^\lambda(x,z) &=\left(
	\begin{array}{c}
		0\\
		\phi^\mathrm{L}_{n_z}(z) \Psi_{\downarrow, \mathrm{L}}^\lambda(x)\\
		0\\
		\phi^\mathrm{H}_{n_z}(z) \Psi_{\downarrow, \mathrm{H}}^\lambda(x)
	\end{array}
	\right),\label{eqn:diagBasis_Bperp21}
\end{align}
where the $z$-part $\phi_{n_z}^{\mathrm{H}/\mathrm{L}}(z)$ is given by Eq.~\eqref{eqn:diagBasis_Bperpz_E} and the spinor components of the in-plane contribution read
\begin{align}
	\Psi_{\uparrow, \mathrm{H}}^\lambda(x) =\  &\psi_{\eta_-+2}^\lambda(L_x/2)c^\lambda_{\uparrow}(\eta_+)\psi_{\eta_+}^\lambda(x)\nonumber\\
	&-\psi_{\eta_++2}^\lambda(L_x/2)c^\lambda_{\uparrow}(\eta_-)\psi_{\eta_-}^\lambda(x), \label{eqn:diagBasis_BperpElem1}\\
	\Psi_{\uparrow, \mathrm{L}}^\lambda(x) =\ &\psi_{\eta_-+2}^\lambda(L_x/2) \psi_{\eta_++2}^\lambda(x)\nonumber\\&-\psi_{\eta_++2}^\lambda(L_x/2)\psi_{\eta_-+2}^\lambda(x),\label{eqn:diagBasis_BperpElem2}\\
	\Psi_{\downarrow, \mathrm{L}}^\lambda(x) =\  &\psi_{\chi_-+2}^\lambda(L_x/2)c^\lambda_{\downarrow}(\chi_+)\psi_{\chi_+}^\lambda(x)\nonumber\\
	&-\psi_{\chi_++2}^\lambda(L_x/2)c^\lambda_{\downarrow}(\chi_-)\psi_{\chi_-}^\lambda(x),\label{eqn:diagBasis_BperpElem3}\\
	\Psi_{\downarrow, \mathrm{H}}^\lambda(x) =\ &\psi_{\chi_-+2}^\lambda(L_x/2) \psi_{\chi_++2}^\lambda(x)\nonumber\\&-\psi_{\chi_++2}^\lambda(L_x/2)\psi_{\chi_-+2}^\lambda(x),\label{eqn:diagBasis_BperpElem4}
\end{align}
where $\eta_{\pm}$ are the two solutions of the quadratic equation
\begin{small}
	\begin{align}
		\frac{\varepsilon_z^{-1/2}(n_z)}{\hbar \omega_c}-\varepsilon+\gamma_{-}\left(\eta+\frac{5}{2}\right)=\frac{3\gamma_s^2 (\eta+2)(\eta+1)\langle\phi_{n_z}^\mathrm{H} |\phi_{n_z}^\mathrm{L}\rangle^2}{\frac{\varepsilon_z^{3/2}(n_z)}{\hbar \omega_c}-\varepsilon+\gamma_{+}\left(\eta+\frac{1}{2}\right)}
	\end{align}
\end{small}
and $\chi_{\pm}$ the solutions of
\begin{small}
	\begin{align}
		\frac{\varepsilon_z^{-3/2}(n_z)}{\hbar \omega_c}-\varepsilon+\gamma_{+}\left(\chi+\frac{5}{2}\right)=\frac{3\gamma_s^2 (\chi+2)(\chi+1)\langle\phi_{n_z}^\mathrm{H} |\phi_{n_z}^\mathrm{L}\rangle^2}{\frac{\varepsilon_z^{1/2}(n_z)}{\hbar \omega_c}-\varepsilon+\gamma_{-}\left(\chi+\frac{1}{2}\right)}.
	\end{align}
\end{small}
The constants are  defined as
\begin{align}
	c_{\uparrow}^e(\eta) &=\frac{(\gamma_1- \gamma_s)(\eta+5/2)-\varepsilon+\varepsilon^{-1/2}_z (n_z)/(\hbar \omega_c)}{\sqrt{3}\gamma_s  \langle \phi^\mathrm{H}_{n_z}|\phi^\mathrm{L}_{n_z}  \rangle (\eta+1)} , \label{eq:parameters_L1}\\
	c_{\uparrow}^o(\eta) &=\frac{(\gamma_1- \gamma_s)(\eta+5/2)-\varepsilon+\varepsilon^{-1/2}_z (n_z)/(\hbar \omega_c)}{\sqrt{3}\gamma_s   \langle \phi^\mathrm{H}_{n_z}|\phi^\mathrm{L}_{n_z}  \rangle (\eta+2)}, \label{eq:parameters_L2}\\
	c_{\downarrow}^e(\chi) &=\frac{(\gamma_1+ \gamma_s)(\chi+5/2)-\varepsilon+\varepsilon^{-3/2}_z (n_z)/(\hbar \omega_c)}{\sqrt{3}\gamma_s  \langle \phi^\mathrm{H}_{n_z}|\phi^\mathrm{L}_{n_z}  \rangle (\chi+1)} , \label{eq:parameters_L1d}\\
	c_{\downarrow}^o(\chi) &=\frac{(\gamma_1+\gamma_s)(\chi+5/2)-\varepsilon+\varepsilon^{-3/2}_z (n_z)/(\hbar \omega_c)}{\sqrt{3}\gamma_s   \langle \phi^\mathrm{H}_{n_z}|\phi^\mathrm{L}_{n_z}  \rangle (\chi+2)}. \label{eq:parameters_L2d}
\end{align}
Imposing the hard-wall boundary conditions on $\Psi_{\downarrow, \mathrm{H}}^\lambda$ and $\Psi_{\downarrow,\mathrm{L}}^\lambda$ we calculate the eigenvalues $\varepsilon$ numerically.

\section{Dispersion relation with negative average mass\label{sec:SpecNegativeMass}}

\begin{figure}[htb]
	\includegraphics{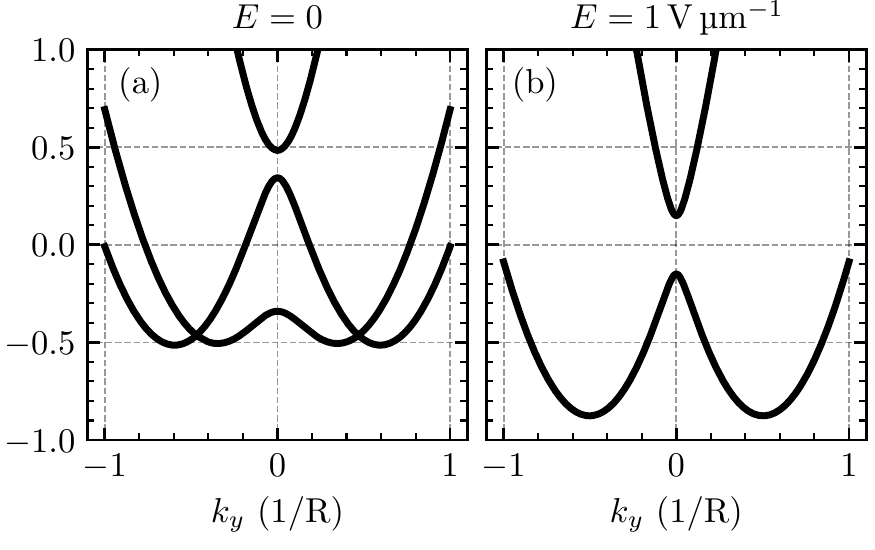}
	\caption{Dispersion relation of the lowest in energy states of a circular NW calculated numerically by diagonalizing the Hamiltonian in Eq.~\eqref{eqn:FullHamiltonian} as described in the main text in Sec~\ref{sec:geff_meff}. In (a) $E=0$ and $\bar{m}<0$, in (b) $E= \SI{1}{\volt\per\micro\meter}$ and $\bar{m}>0$. Here, $B = \SI{2}{\tesla}, R= \SI{22}{\nano\meter}/\sqrt{\pi}$.\label{fig:dispRelCirc}}
\end{figure}

In this appendix we show a plot of the dispersion relation of a NW with circular cross-section of radius $R= \SI{22}{\nano\meter}/\sqrt{\pi}$. We choose the same parameters as in Fig.~\ref{fig:effPar_R22_rect-circ} in the main text and show in Fig.~\ref{fig:dispRelCirc}(a) the dispersion relation with negative average effective mass and in Fig.~\ref{fig:dispRelCirc}(b) with positive effective mass. Interestingly, we observe a crossing between the lowest two states close to $k_y = 1/(2R)$ at $E=0$ which becomes an anti-crossing for larger electric field due to the SOI. As already mentioned in the main text, terms higher order in $k_y$  make sure that the dispersion relation has a positive curvature at large values of $k_y$ even when $\bar{m}<0$.

\begin{widetext}
	
\section{Rotations of the LK Hamiltonian\label{sec:rotations}}

\renewcommand{\arraystretch}{1.5}
\begin{table*}[htbp!]
	\caption{For each line in Fig.~\ref{fig:growthDir} we fix one axis and then rotate around this axis as shown below. We use the coordinate system with $\mathrm{NW} \parallel z$ and $\vect{B} \parallel x$. \label{tab:growthDirections}}
	\begin{ruledtabular}
		\begin{tabular}{c|cc}
			fixed axis& & \\\hline\hline
			$z\parallel [001]$ & $x\parallel \cos(\varphi)  [100] - \sin (\varphi)  [010]$ & $y\parallel \sin(\varphi)  [100]  +\cos (\varphi)  [010]$\\\hline
			$x\parallel [001]$ & $y\parallel \sin(\varphi)  [100] + \cos (\varphi)  [010]$ & $z\parallel -\cos(\varphi)  [100] + \sin(\varphi)  [010]$\\\hline
			$z\parallel [110]$ &  $x\parallel \frac{\cos(\varphi) }{\sqrt{2}} [100] - \frac{\cos(\varphi) }{\sqrt{2}} [010] + \sin (\varphi)  [001]$ & $y\parallel \frac{\sin(\varphi) }{\sqrt{2}}  [100] -\frac{\sin(\varphi) }{\sqrt{2}} [010] - \cos(\varphi)  [001]$\\\hline
			$z\parallel [111]$ & $\begin{array}{@{}c@{}}
				x\parallel \left(\frac{\cos(\varphi)}{\sqrt{6}}+\frac{\sin(\varphi)}{\sqrt{2}}\right)[100] + \left(\frac{\cos(\varphi)}{\sqrt{6}}-\frac{\sin(\varphi)}{\sqrt{2}} \right)[010]\\- \sqrt{\frac{2}{3}} \cos(\varphi) [001]
			\end{array}$
			& $\begin{array}{@{}c@{}}y\parallel \left(-\frac{\cos(\varphi)}{\sqrt{2}}+\frac{\sin(\varphi)}{\sqrt{6}}\right)[100] + \left(\frac{\cos(\varphi)}{\sqrt{2}}+\frac{\sin(\varphi)}{\sqrt{6}} \right)[010]\\- \sqrt{\frac{2}{3}} \sin(\varphi) [001]\end{array}$ 
		\end{tabular}
	\end{ruledtabular}
\end{table*}

In the main text we consider the situations where NW$ \parallel y$ as well as $z \parallel \vect{B}$. In the following we switch to a different coordinate system in order to express the LK Hamiltonian where the spin quantization axis is aligned along the NW axis. This is a more convenient basis for the interpretation of the matrix elements of the LK Hamiltonian. To switch to the new coordinate system, we replace $x \rightarrow y,\ y\rightarrow z,\ z \rightarrow x$. Then the NW is parallel to the $z$-axis and the magnetic field is parallel to the $x$-axis.

In order to make $x\parallel [001]$, we rotate the LK Hamiltonian by $\pi/2$ around the $[010]$-axis and keep this axis fixed. Then, $z\parallel [110]$ is obtained in our case by a rotation of  $\pi/2$ around the $[001]$-axis followed by $\pi/4$ around $[001]$ which gives $x\parallel [1\bar{1}0]$ and $y\parallel [00\bar{1}]$. The last case of $z\parallel [111]$ comes from a rotation first of $-\arctan (\sqrt{3})$ around the $[010]$-axis followed by a $-\pi/4$-rotation around the $[001]$-axis.  We keep these axes fixed for each case and then rotate by the angle $\varphi$ around this axis according to Tab.~\ref{tab:growthDirections} and as illustrated in the top of Fig.~\ref{fig:growthDir}.

The rotations around the fixed main crystallographic axes are performed via standard Euler rotation matrices  $\mathcal{R}$.
Then, we only need to solve the following equations for the momenta $k_j$ and spin-$3/2$ matrices $J_j$, $j = x, y, z$, 
\begin{align}
	\begin{pmatrix}
		k_{x'}\\ k_{y' }\\ k_{z'}
	\end{pmatrix} = \mathcal{R} \begin{pmatrix}
	k_x\\ k_y\\ k_z
\end{pmatrix}, \ 
\begin{pmatrix}
	J_{x'}\\ J_{y' }\\ J_{z'}
\end{pmatrix} = \mathcal{R} \begin{pmatrix}
	J_x\\ J_y\\ J_z
\end{pmatrix}
\end{align}
and plug them into the LK Hamiltonian in Eq.~\eqref{eqn:LK_Hamiltonian}. For certain angles $\varphi$ we give the crystallographic directions to which the coordinate axes are parallel in Tab.~\ref{tab:growthDirectionsExplicit}.
\begin{table*}[htbp!]
	\caption{Explicit crystallographic directions for certain rotation angles $\varphi$. We use the coordinate system with $\mathrm{NW} \parallel z$ and $\vect{B} \parallel x$. \label{tab:growthDirectionsExplicit}}
	\begin{ruledtabular}
		\begin{tabular}{c|ccccc}
			& $\varphi=0$ & $\varphi=\pi/4$ & $\varphi=\pi/2$ & $\varphi=3\pi/4$ & $\varphi=\pi$\\\hline\hline
			$z\parallel [001]$ & $x\parallel[100], y\parallel[010]$ & $x\parallel[1\bar{1}0], y\parallel[110]$ & $x\parallel[0\bar{1}0], y\parallel[100]$ & $x\parallel[\bar{1}\bar{1}0],y\parallel[1\bar{1}0]$ & $x\parallel[\bar{1}00], y\parallel[0\bar{1}0]$ \\\hline
			$x\parallel [001]$ & $y\parallel[010], z\parallel[\bar{1}00]$ & $y\parallel[110], z\parallel[\bar{1}10]$ & $y\parallel[100], z\parallel[010]$ & $y\parallel[1\bar{1}0], z\parallel[110]$ & $y\parallel[0\bar{1}0], z\parallel[100]$\\\hline
			$z\parallel [110]$ &  $x\parallel[1\bar{1}0], y\parallel[00\bar{1}]$ &  & $x\parallel[001], y\parallel[1\bar{1}0]$ & & $x\parallel[\bar{1}10], y\parallel[001]$\\\hline
			$z\parallel [111]$ & $x\parallel[11\bar{2}], y\parallel[\bar{1}10]$ &  & $x\parallel[1\bar{1}0], y\parallel[11\bar{2}]$ & & $x\parallel[\bar{1}\bar{1}2], y\parallel[1\bar{1}0]$ 
		\end{tabular}
	\end{ruledtabular}
\end{table*}

The LK Hamiltonians for the different growth directions fixed along the $z$-axis discussed in the main text are, given in the coordinate system with $\mathrm{NW} \parallel z$ and $\vect{B} \parallel x$,
	\begin{align}
		H_\mathrm{LK}^{z\parallel[001]} = 
		\frac{\hbar^2}{2 m} & \left[
		\begin{pmatrix}
			(\gamma_1 + \gamma_2) \pi_+ \pi_-  & 0 & M & 0\\
			0 & (\gamma_1 - \gamma_2) \pi_+ \pi_- & 0 & M\\
			M^\ast & 0 & (\gamma_1 - \gamma_2) \pi_+ \pi_- & 0 \\
			0 & M^\ast & 0 & (\gamma_1 + \gamma_2) \pi_+ \pi_-
		\end{pmatrix} \right.\nonumber\\
		+& \begin{pmatrix}
			0 & - 2\sqrt{3} \gamma_3 \pi_- & 0 & 0 \\
			- 2\sqrt{3} \gamma_3 \pi_+ & 0 & 0 & 0 \\
			0 & 0 & 0 & 2\sqrt{3} \gamma_3 \pi_-\\
			0 & 0 & 2\sqrt{3} \gamma_3 \pi_+ & 0 
		\end{pmatrix} \pi_z
		+  \left. \begin{pmatrix}
			\gamma_1 - 2 \gamma_2 & 0 & 0 & 0\\
			0 &  \gamma_1 + 2 \gamma_2 & 0 & 0 \\
			0 & 0 & \gamma_1 + 2 \gamma_2 & 0 \\
			0 & 0 & 0 & \gamma_1 - 2 \gamma_2
		\end{pmatrix} \pi_z^2 \right], \label{eqn:LKHam_z001}
\\
			H_\mathrm{LK}^{z\parallel[110]} = 
\frac{\hbar^2}{2 m} & \left[
\begin{pmatrix}
	N_+ & 0 & O & 0 \\
	0 & N_-  & 0 & O\\
	O^\ast& 0 & N_- & 0\\
	0 &  O^\ast & 0 &  N_+
\end{pmatrix} \right.
+ \begin{pmatrix}
	0 & -P & 0 & 0\\
	-P^\ast& 0 & 0 & 0 \\
	0 & 0 & 0 & P\\
	0 & 0 &P^\ast & 0 
\end{pmatrix} \pi_z\nonumber\\
+& \left.\begin{pmatrix}
	\frac{1}{2}(2\gamma_1-\gamma_2 -3 \gamma_3) & 0 & \frac{\sqrt{3}}{2} e^{- 2 i \varphi}(\gamma_3- \gamma_2) & 0 \\
	0 & \frac{1}{2} (2\gamma_1+\gamma_2 +3 \gamma_3) & 0 & \frac{\sqrt{3}}{2}e^{- 2 i \varphi}(\gamma_3- \gamma_2)\\
	\frac{\sqrt{3}}{2}e^{2 i \varphi}(\gamma_3- \gamma_2) & 0 & \frac{1}{2} (2\gamma_1+\gamma_2 +3 \gamma_3) & 0 \\
	0 & \frac{\sqrt{3}}{2}e^{2 i \varphi}(\gamma_3- \gamma_2) & 0 & \frac{1}{2}(2\gamma_1-\gamma_2 -3 \gamma_3)
\end{pmatrix} \pi_z^2 \right],\label{eqn:LKHam_z110}
\end{align}
\begin{align}
		H_\mathrm{LK}^{z\parallel[111]} = 
		\frac{\hbar^2}{2m} &\left[
		\begin{pmatrix}
			(\gamma_1+\gamma_3)\pi_+\pi_- & \sqrt{\frac{2}{3}} e^{-3 i \varphi} (\gamma_2-\gamma_3) \pi_+^2 & - \frac{\gamma_2 + 2 \gamma_3}{\sqrt{3}} \pi_-^2 & 0 \\
			\sqrt{\frac{2}{3}} e^{3 i \varphi} (\gamma_2-\gamma_3) \pi_-^2 &  (\gamma_1-\gamma_3)\pi_+ \pi_- & 0 & - \frac{\gamma_2 + 2 \gamma_3}{\sqrt{3}} \pi_-^2  \\
			- \frac{\gamma_2 + 2 \gamma_3}{\sqrt{3}} \pi_+^2 & 0 & 	(\gamma_1-\gamma_3)\pi_+ \pi_- & -\sqrt{\frac{2}{3}} e^{-3 i \varphi} (\gamma_2-\gamma_3) \pi_+^2 \\
			0 & - \frac{\gamma_2 + 2 \gamma_3}{\sqrt{3}} \pi_+^2 & -\sqrt{\frac{2}{3}} e^{3 i \varphi} (\gamma_2-\gamma_3) \pi_-^2  & 	(\gamma_1+\gamma_3)\pi_+ \pi_-
		\end{pmatrix} \right.\nonumber\\
	+& \begin{pmatrix}
		0 & -\frac{2}{\sqrt{3}} (2 \gamma_2 + \gamma_3) \pi_- & \sqrt{\frac{8}{3}} e^{-3 i \varphi} (\gamma_2 - \gamma_3) \pi_+ & 0\\
		  -\frac{2}{\sqrt{3}} (2 \gamma_2 + \gamma_3) \pi_+ & 0 & 0 & \sqrt{\frac{8}{3}} e^{-3 i \varphi} (\gamma_2 - \gamma_3) \pi_+\\
		 \sqrt{\frac{8}{3}} e^{3 i \varphi} (\gamma_2 - \gamma_3) \pi_- & 0 & 0 & \frac{2}{\sqrt{3}} (2 \gamma_2 + \gamma_3) \pi_- \\
		  0 & \sqrt{\frac{8}{3}} e^{3 i \varphi} (\gamma_2 - \gamma_3) \pi_- & \frac{2}{\sqrt{3}} (2 \gamma_2 + \gamma_3) \pi_+ & 0
	\end{pmatrix} \pi_z  \nonumber\\
 + &\left. \begin{pmatrix}
 	\gamma_1-2 \gamma_3 & 0 & 0 & 0\\
 	0 & \gamma_1 + 2\gamma_3 & 0 & 0 \\
 	0 & 0 & \gamma_1 + 2 \gamma_3 & 0\\
 	0 & 0 & 0 & \gamma_1- 2\gamma_ 3
 \end{pmatrix} \pi_z^2\right].\label{eqn:LKHam_z111}
	\end{align}
with $\pi_\pm = \pi_x \pm i \pi_y$. The matrix elements are defined as
\begin{align}
	M =&  -\frac{\sqrt{3}}{2}  \left(e^{-4 i \varphi} (\gamma_2 - \gamma_3)\pi_+^2\ +  (\gamma_2 +\gamma_3) \pi_-^2 \right),  \label{eqn:matElemM} \\
	N_\pm =& \frac{1}{8}\left[\mp 3 (\gamma_2-\gamma_3) (e^{- 2 i \varphi} \pi_+ + e^{2 i \varphi} \pi_-) + 2(4\gamma_1 \pm \gamma_2 \pm 3 \gamma_3) \pi_+ \pi_-\right],\\
	O =& \frac{\sqrt{3}}{8}\left(3  e^{-4 i \varphi} (\gamma_3-\gamma_2)\pi_+^2 + 2 e^{-2 i \varphi} (\gamma_2-\gamma_3) \pi_+ \pi_- - (3\gamma_2 + 5 \gamma_3) \pi_-^2 \right), \label{eqn:matElemO}\\
	P =& \sqrt{3} \left(e^{- 2 i \varphi} (\gamma_2 -\gamma_3) \pi_++ (\gamma_2 + \gamma_3) \pi_m \right).
\end{align}
Looking at the $\pi_z = k_z=0$ parts of the Hamiltonians explains the periodicity of the $g$-factor under rotation by the angle $\varphi$ around the labeled fixed axis. The off-diagonal matrix elements $M$, cf. Eq.~\eqref{eqn:matElemM}, of the Hamiltonian in Eq.~\eqref{eqn:LKHam_z001} at $k_z = 0$ contain the exponential $\exp(-4 i \varphi)$ and couple the spins $\pm 3/2$ and $\mp 1/2$. This explains the $\pi/2$-periodicity of the effective $g$-factor for $\mathrm{NW}\parallel [001]$. Similarly, the Hamiltonian in Eq.~\eqref{eqn:LKHam_z110} at $k_z=0$ contains off-diagonal matrix elements $O$ which have a term $\propto \exp(-2 i \varphi)$, cf. Eq.~\eqref{eqn:matElemO}. Also, these matrix elements couple spin $\pm 3/2$ to spin $\mp 1/2$ and, thus, explain the $\pi$-periodicity of the $g$-factor for $\mathrm{NW}\parallel [110]$. In contrast to that, the Hamiltonian in Eq.~\eqref{eqn:LKHam_z111} at $k_z=0$ contains a $\varphi$-dependence only in the matrix elements coupling spin $\pm 3/2$ and $\pm 1/2$. Hence, only the phase of the HH-LH wave function is changed and not the amplitude leaving the $g$-factor for $\mathrm{NW}\parallel [111]$ unchanged. 

The LK Hamiltonian with $x \parallel [001]$, again in the coordinate system with $\mathrm{NW} \parallel z$ and $\vect{B} \parallel x$, reads
\begin{align}
	H_\mathrm{LK}^{x\parallel[001]} = 
	\frac{\hbar^2}{2 m} & \left[
	\begin{pmatrix}
		Q_+& \frac{i \sqrt{3}}{2} (\gamma_2 - \gamma_3) \sin(4 \varphi) \pi_y^2& R & 0\\
		-\frac{i \sqrt{3}}{2} (\gamma_2 - \gamma_3) \sin(4 \varphi) \pi_y^2& Q_- & 0 & R\\
		R^\ast & 0 & Q_-& \frac{i \sqrt{3}}{2} (\gamma_2 - \gamma_3) \sin(4 \varphi) \pi_y^2 \\
		0 & R^\ast & -\frac{i \sqrt{3}}{2} (\gamma_2 - \gamma_3) \sin(4 \varphi) \pi_y^2 & Q_+
	\end{pmatrix} \right.\nonumber\\
	+& \begin{pmatrix}
		\frac{3}{2}(\gamma_2 - \gamma_3) \sin(4\varphi) \pi_y & S & \frac{\sqrt{3}}{2}(\gamma_2-\gamma_3) \sin(4\varphi) \pi_y & 0 \\
		S^\ast & -\frac{3}{2}(\gamma_2 - \gamma_3) \sin(4\varphi) \pi_y & 0 & \frac{\sqrt{3}}{2}(\gamma_2-\gamma_3) \sin(4\varphi) \pi_y \\
		\frac{\sqrt{3}}{2}(\gamma_2-\gamma_3) \sin(4\varphi) \pi_y& 0 & -\frac{3}{2}(\gamma_2 - \gamma_3) \sin(4\varphi) \pi_y & -S\\
		0 & \frac{\sqrt{3}}{2}(\gamma_2-\gamma_3) \sin(4\varphi) \pi_y& -S^\ast& \frac{3}{2}(\gamma_2 - \gamma_3) \sin(4\varphi) \pi_y 
	\end{pmatrix} \pi_z\nonumber\\
	+&  \left. \begin{pmatrix}
		T_+ & -\frac{i \sqrt{3}}{2}(\gamma_2 - \gamma_3) \sin(4\varphi) & \frac{\sqrt{3 }}{2} (\gamma_2-\gamma_3) \sin^2(2\varphi) & 0 \\
		\frac{i \sqrt{3}}{2}(\gamma_2 - \gamma_3) \sin(4\varphi) & T_- & 0 & \frac{\sqrt{3 }}{2} (\gamma_2-\gamma_3) \sin^2(2\varphi) \\
		\frac{\sqrt{3 }}{2} (\gamma_2-\gamma_3) \sin^2(2\varphi)  & 0 & T_- & \frac{i \sqrt{3}}{2}(\gamma_2 - \gamma_3) \sin(4\varphi)\\
		0 & \frac{\sqrt{3 }}{2} (\gamma_2-\gamma_3) \sin^2(2\varphi) & -\frac{i \sqrt{3}}{2}(\gamma_2 - \gamma_3) \sin(4\varphi) & T_+
	\end{pmatrix} \pi_z^2 \right], \label{eqn:LKHam_x001}
\end{align}
with the explicit matrix elements
\begin{align}
	Q_\pm =& \frac{1}{4}\left[4 (\gamma_1\pm\gamma_2) \pi_x^2+ (4\gamma_1 \pm \gamma_2 \pm 3 \gamma_3) \pi_y^2 \pm 3 (\gamma_2- \gamma_3) \cos(4 \varphi)\right],\\
	R =& \frac{\sqrt{3}}{4} \left[-4 \gamma_2 \pi_x^2 + 8 i \gamma_3 \pi_x \pi_y + (3\gamma_2 + \gamma_3) \pi_y^2+(\gamma_2 -\gamma_3) \cos(4\varphi) \pi_y^2 \right], \label{eqn:matElemR}\\
	S =& i \sqrt{3} [ 2 i \gamma_3 \pi_x + (\gamma_2 + \gamma_3) \pi_y - (\gamma_2-\gamma_3) \cos(4 \varphi) \pi_y ],\\
	T_\pm =& \frac{1}{4}\left[4 \gamma_1 \mp 5 \gamma_2 \mp3 \gamma_3 \mp 3 (\gamma_2- \gamma_3) \cos(4 \varphi)\right].
\end{align}
The $g$-factor for $\vect{B} \parallel [001]$ is again $\pi/2$-periodic which is explained well by the matrix element $R$ in Eq.~\eqref{eqn:matElemR}, the only matrix element coupling spins $\pm 3/2$ and $\mp 1/2$ at $k_z = 0$. The matrix element $R$ contains a term $\propto \cos(4 \varphi)$ leading to the $\pi/2$-periodicity. 
\end{widetext}

\bibliography{Literature.bib}

\end{document}